\documentclass[%
 12pt,
]{iopart}

\expandafter\let\csname equation*\endcsname\relax
\expandafter\let\csname endequation*\endcsname\relax
\usepackage{amsmath}

\usepackage{amssymb}
\usepackage{graphicx}
\usepackage{bm}
\usepackage{braket}
\usepackage{booktabs}
\usepackage{color}
\usepackage{siunitx}
\usepackage{multirow}

\newcommand{\mrm}[1]{\mathrm{#1}}

\begin{document}
\title[Predicting excited states from ground state wavefunction]{Predicting excited states from ground state wavefunction by supervised quantum machine learning}

\author{Hiroki Kawai$^1$ and Yuya O. Nakagawa$^2$}
\address{$^1$
Department of Physics, 590 Commonwealth Avenue
Boston, MA, United States, 02215\\
Electrical and Computer Engineering
Department, Boston University, 8 Saint Mary's Street, Boston, MA, United States, 02215}
\ead{hirokik@bu.edu}
\bigskip
\address{$^2$
QunaSys Inc., Aqua Hakusan Building 9F, 1-13-7 Hakusan, Bunkyo, Tokyo 113-0001, Japan}
\ead{nakagawa@qunasys.com}

\vspace{10pt}
\begin{indented}
\item[] July 2020
\end{indented}

\begin{abstract}
Excited states of molecules lie in the heart of photochemistry and chemical reactions.
The recent development in quantum computational chemistry leads to inventions of a variety of algorithms that calculate the excited states of molecules on near-term quantum computers, but they require more computational burdens than the algorithms for calculating the ground states.
In this study, we propose a scheme of supervised quantum machine learning which predicts the excited-state properties of molecules only from their ground state wavefunction resulting in reducing the computational cost for calculating the excited states.
Our model is comprised of a quantum reservoir and a classical machine learning unit which processes the measurement results of single-qubit Pauli operators with the output state from the reservoir.
The quantum reservoir effectively transforms the single-qubit operators into complicated multi-qubit ones which contain essential information of the system, so that the classical machine learning unit may decode them appropriately.
The number of runs for quantum computers is saved by training only the classical machine learning unit,
and the whole model requires modest resources of quantum hardware that may be implemented in current experiments.
We illustrate the predictive ability of our model by numerical simulations for small molecules with and without noise inevitable in near-term quantum computers.
The results show that our scheme well reproduces the first and second excitation energies as well as the transition dipole moment between the ground states and excited states only from the ground state as an input. 
We expect our contribution will enhance the applications of quantum computers in the study of quantum chemistry and quantum materials.
\end{abstract}

\maketitle
\section{Introduction \label{sec:intro}}
The rapid growth of the machine learning technology in the last decade has revealed its potential to be utilized in various engineering fields such as image recognition, natural language processing, and outlier detection~\cite{Lecun2015, Goodfellow2016}.
Its applications to scientific fields have also attracted numerous attentions recently as well as those to engineering.
One of the most active research areas is physical science~\cite{Carleo2019}, especially studies of quantum many-body systems including condensed matter physics and quantum chemistry.
For example, one can classify a phase of matter from its wavefunction~\cite{Carrasquilla2017, Nieuwenburg2017} or predict the atomization energy of molecules~\cite{Rupp2012, Montavon2012, Hansen2013} from their molecular structures with sophisticated machine learning techniques.

Most of those researches employ {\it classical} machine learning, with which classical data are processed by classical algorithms and computers.
On the other hand, machine learning algorithms on a quantum processor have been developed since the invention of the Harrow-Hassidim-Lloyd (HHL) algorithm, and they are dubbed as ``quantum machine learning" ~\cite{Harrow2009, Wiebe2012, Lloyd2014, Rebentrost2014, Schuld2016, Kerenidis2020, Wang2017, Biamonte2017}. 
In the last few years, there has been surging interest in quantum machine learning leveraging the variational method ~\cite{Cong2019, Havlicek2019, Kusumoto2019, Mitarai2018, Farhi2018, Wilson2018}, in which a shallow quantum circuit parameterized with classical parameters such as the angles of the rotational gates is optimized with a classical optimization algorithm to find optimal parameters for performing the given objective. This is because a primitive type of quantum computers is about to be realized in the near future, and such machines may have the potentials to outperform classical computers ~\cite{Harrow2017, Arute2019}. 
Those near-term quantum computers are called noisy intermediate-scale quantum (NISQ) devices ~\cite{Preskill2018} and consist of hundreds to thousands of physical, non-fault-tolerant qubits.

So far, quantum machine learning has been mostly applied to classical computing tasks with classical data such as pattern recognition of images~\cite{Havlicek2019, Kusumoto2019, Wilson2018}.
In those studies, the classical data must be encoded in quantum states to be processed by quantum computers, but the encoding is generally inefficient; it requires the exponentially large number of gates to encode classical data into a quantum state unless the data have a structure of the tensor product which is compatible with that of qubits~\cite{Giovannetti2008PRL, Giovannetti2008PRA, Prakash2014}.

Therefore, it is natural to think of performing tasks with {\it quantum nature}.
In this study, we consider the following task: predicting excited-state properties of a given molecular system from its ground state wavefunction. 
Specifically, we are interested in the Hamiltonian for the electronic states of molecules. 
The question we raise and want to solve leveraging quantum machine learning 
is whether it is possible to predict properties of the excited states 
from the ground state wavefunction $\ket{\psi_0}$.
According to the celebrated Hohenberg-Kohn theorem~\cite{Hohenberg1964}, one can determine an external potential for electrons and thereby the whole original electron Hamiltonian from its ground state electron density $\rho_0(r) = \braket{\psi_0|\hat{r}|\psi_0}$ up to constant, where $\hat{r}$ is the position operator.
Hence, it should be also possible to predict the excited states from the ground state in principle.

The task we propose here has various practical and conceptual attractions from the viewpoint of quantum machine learning and the studies of quantum many-body systems.
First, practically, computing excited states of a given Hamiltonian needs significantly larger computational cost and is more difficult than computing the ground state~\cite{Roos2012, Helgaker2014}.
Since the excited-state properties are essential for thermodynamics of the system and non-equilibrium dynamics such as chemical reactions, a large benefit to the studies of quantum chemistry and quantum materials is expected if one may predict the excited states only from the ground state.  
We note that applying classical machine learning to predict excited states of molecules from classical data (molecular structure, coulomb matrix, etc.) has been widely explored in the literature~\cite{Goh2017, Montavon2013, Ramakrishnan2015, Hase2016, Wu2018}.
Second, the problem of encoding data to quantum computers mentioned above can be circumvented in this setup; as we will see later, it is possible to input wavefunctions into quantum registers directly from outputs of another quantum algorithm which yields a ground state wavefunction, such as the variational quantum eigensolver (VQE), which is one of the most promising applications of the NISQ devices ~\cite{Peruzzo2014}.
Third, from a conceptual point of view, the original ``data" of quantum systems are wavefunctions, which are quantum in nature, so quantum machine learning dealing with quantum data {\it as they are} will take advantage of the whole information contained in the wavefunctions and potentially has stronger predictability than the classical counterparts which process only classical features of quantum data in a pure classical way~\cite{Sasaki2001, Sasaki2002}.

In this study, we propose a simple quantum machine learning scheme to predict the excited-state properties of the Hamiltonian of a given molecule from its ground state wavefunction.
Our simulations suggest the potential that one can implement our model on the real NISQ devices being robust to the inevitable noise of outputs on such devices.
In particular, we employ and generalize the quantum reservoir computing~\cite{Fujii2017} and quantum reservoir processing~\cite{Ghosh2019} techniques. 

Both techniques feed the initial quantum information to a random quantum system called a ``quantum reservoir" which evolves the initial state to another state, and the measurement results of the output state are learned by linear regression to predict some properties associated with the initial information. 

Similarly, we first process an input wavefunction that is the ground state of the target molecular Hamiltonian with a random quantum circuit or the time evolution under another certain Hamiltonian and then measure the expectation values of one-qubit operators afterwards.

The measurement results are post-processed by a classical machine learning unit, and we train only the classical unit to predict the target properties of the system by supervised learning so that the overall number of runs of quantum computers is small.
In the Heisenberg picture, the quantum reservoir effectively transforms the one-qubit operators into complicated multi-qubit ones which contain essential information of the system, and the classical machine learning will decode them appropriately. 
We numerically demonstrate the predictive power of our scheme by taking three small molecules as examples.
Our model can predict the excitation energies and the transition dipole moment between the ground state and the excited state properly only from the ground state wavefunction.

The rest of the paper is organized as follows.
In Sec.~\ref{sec:method}, we explain our setup in detail and propose a model for quantum machine learning of excited states, besides presenting the way to train the model.
In Sec.~\ref{sec:result}, we show the result of numerical simulations of our scheme predicting the excited-state properties of small molecules as examples. 
Section~\ref{sec:discussion} is dedicated for the discussion of our result.
We conclude the study in Sec.~\ref{sec:conclusion}.
\ref{apdx:vqe} is a review of the VQE and its extension to find the excited states.
\ref{apdx:jw-transformation} introduces the Jordan-Wigner transformation, which is used to map fermionic molecular Hamiltonians into Hamiltonians written in qubit operators. 
\ref{apdx:non-linearity} is the extension of the discussion in Section~\ref{sec:discussion} to demonstrate the non-linearity between the excited-state properties of the Hamiltonian and the information one may obtain from the ground state. 
\ref{apdx:entangler-analysis} provides a further analysis of the effect of the entangler. 
The dependence of our scheme on the performance of the VQE is analyzed in \ref{apdx:vqe-shots}.

\section{Method \label{sec:method}}
\begin{figure}
\includegraphics[width=1.0\linewidth]{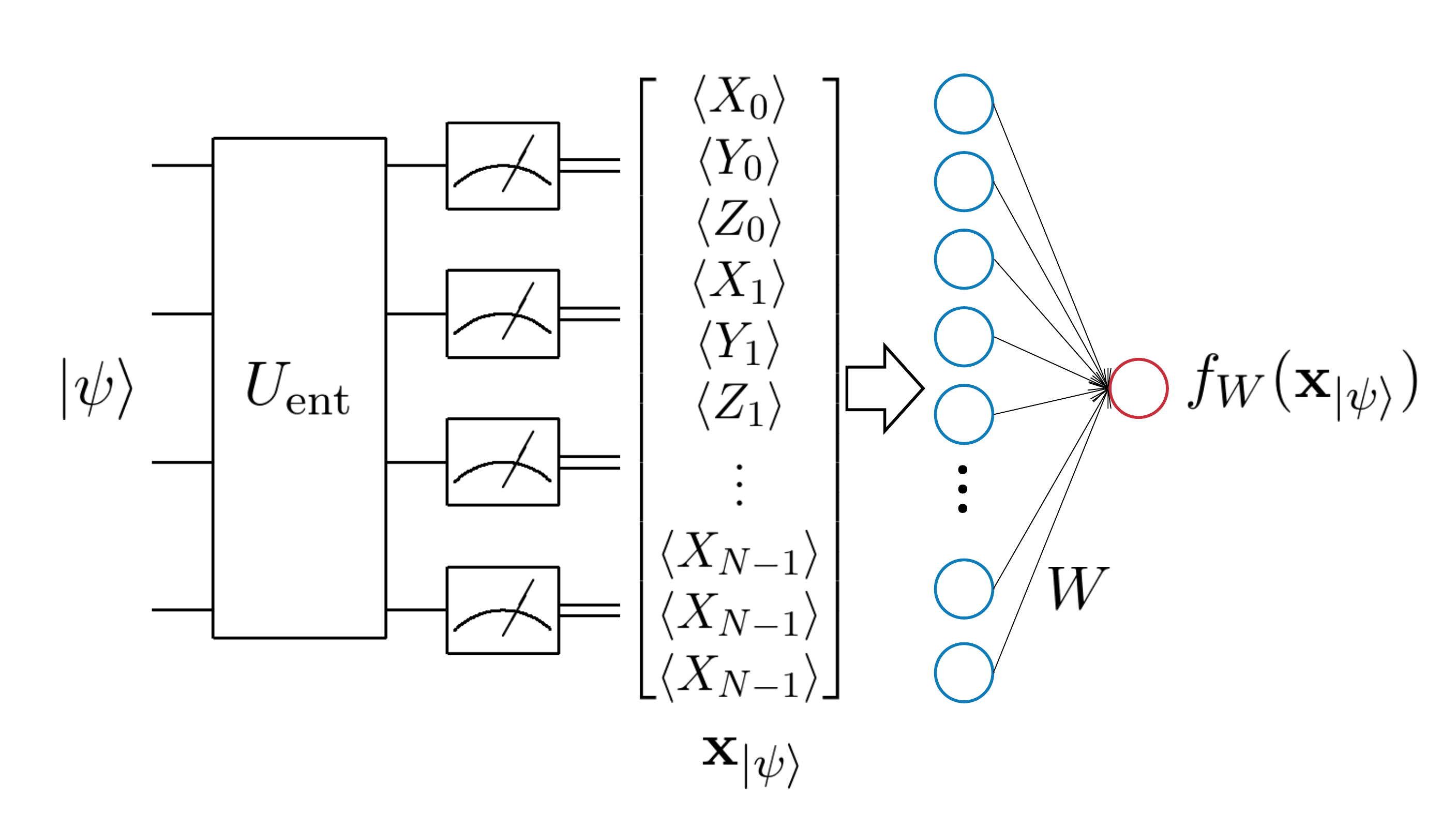}
\caption{Schematic diagram of our model for quantum machine learning of the excited-state properties of a molecule from its ground state.

The input qubit state $\ket{\psi}$, which is assumed to be the ground state, is processed with a quantum circuit $U_\mrm{ent}$, and the measurement yields a classical vector $\mathbf{x}_{\ket{\psi}}$, whose elements $\braket{X_i}, \braket{Y_i}, \braket{Z_i}$ are the expectation values of the single-qubit measurements of the Pauli operators $X, Y, Z$, respectively, on the $i$-th qubit.

A classical machine learning unit $f_W$ with learnable parameters $W$ outputs the target properties from $\mathbf{x}_{\ket{\psi}}$.} \label{fig:model_diagram}
\end{figure}

In this section, we propose a model for quantum machine learning and explain its training process.
The schematic diagram of our model is described in Figure~\ref{fig:model_diagram}.

\subsection{Model description\label{sec:model-desc}}
Let us consider an $N$-qubit system and a wavefunction $\ket{\psi} \in \mathbb{C}^{2^N}$ on it.
Our learning model proceeds as follows.
First, an input $N$-qubit state $\ket{\psi}$, which is assumed to be the ground state of a given Hamiltonian here, is prepared on a quantum computer and
fed into a quantum circuit which is denoted as $U_\mrm{ent}$ in Figure~\ref{fig:model_diagram}.
We call this circuit a quantum entangler or a quantum reservoir for its role of mixing local quantum information of the input state $\ket{\psi}$ and encoding it to the output state $U_\mrm{ent}\ket{\psi}$.
$U_\mrm{ent}$ is chosen to create enough entanglement in the wavefunction and fixed for each learning task (or an experiment).
The details of $U_\mrm{ent}$ are not so important for the quality of learning as illustrated by an exactly-solvable model in Sec.~\ref{sec:discussion},
so one can use a quantum circuit easy to be realized on real quantum devices.
After applying $U_\mrm{ent}$, we measure the expectation values of local Pauli operators $\{X_0, Y_0, Z_0, \cdots, X_{N-1}, Y_{N-1}, Z_{N-1}\}$, where $X_i$, $Y_i$, and $Z_i$ represent a Pauli $X, Y, Z$ operator acting on the site $i$, respectively.
Although the total number of operators is $3N$, we can measure the operators $X_0, ..., X_{N-1}$ simultaneously since they commute with each other, so can we for the cases of $Y_0, ..., Y_{N-1}$ and $Z_0, ..., Z_{N-1}$.

Hence, one can measure all operators with only three different circuits, i.e., the number of experiments to obtain the measurement data does not scale with the number of qubit $N$, but only with the desired precision $\epsilon$ as $\mathcal{O}(1/\epsilon^2)$ due to the statistical uncertainty.

After the measurements, we obtain a $3N$-dimensional real-valued classical vector:
\begin{align}\label{eq:x-vec}
\mathbf{x}_{\ket{\psi}}
= \left( \braket{X_0}, \cdots, \braket{Z_{N-1}} \right)^T 
= \left( \braket{\psi|U_\mrm{ent}^\dagger X_0 U_\mrm{ent}|\psi}, \cdots, \braket{\psi|U_\mrm{ent}^\dagger Z_{N-1} U_\mrm{ent}|\psi} \right)^T.    
\end{align}
Finally, the classical data $\mathbf{x}_{\ket{\psi}}$ is fed into a classical machine learning unit with learnable parameters $W$, such as a linear regression model or a neural network, and the prediction $f_W(\mathbf{x}_{\ket{\psi}})$ is obtained.

We have several comments in order.
First, the process to obtain $\mathbf{x}_{\ket{\psi}}$ from $\ket{\psi}$ can be viewed as compressing the data of $2^N$-dimensional complex-valued vector $\ket{\psi}$ into $3N$-dimensional real-valued data.
Although the way of compression is quite complicated due to the entangler $U_\mrm{ent}$,
the classical machine learning unit can decode the information in $\mathbf{x}_{\ket{\psi}}$ and use it to predict the properties of the excited states of the Hamiltonian.

More concretely, the effect of the entangler is to make the classical vector $\mathbf{x}_{\ket{\psi}}$ to contain the expectation values of complicated (generally long-ranged, many-body) observables for the original ground state $\ket{\psi}$; that is, $\mathbf{x}_{\ket{\psi}}$ can be viewed as the expectation values of the complicated operators $\{U_\mrm{ent}^\dag X_i U_\mrm{ent}, U_\mrm{ent}^\dag Y_i U_\mrm{ent}, U_\mrm{ent}^\dag Z_i U_\mrm{ent}\}_{i=0}^{N-1}$ for the ground state $\ket{\psi}$.
When we expand $U_\mrm{ent}^\dag X_i U_\mrm{ent}$  as $U_\mrm{ent}^\dag X_i U_\mrm{ent} = \sum_j \lambda_j^{(X_i)} P_j^{(X_i)}$, where $\lambda_j^{(X_i)}$ is some coefficient and $P_j^{(X_i)}$ is an $N$-qubit Pauli operator, some of $\{P_j^{(X_i)}\}_j$ are the long-ranged and many-body ones if $U_\mrm{ent}$ creates entanglement over the whole system.
This means that the classical vector  $\mathbf{x}_{\ket{\psi}}$ contains a lot of detailed information of $\ket{\psi}$ as multi-point, long-ranged correlation functions even though we measure only the single-qubit Pauli operators $\bigcup_{i=0}^{N-1}\{X_i, Y_i, Z_i\}$ in reality.
Although how such information is implemented in $\mathbf{x}_{\ket{\psi}}$ is not explicitly known since we do not know the actual values of coefficients $\lambda_j^{(X_i)}$, the classical machine learning unit can be trained to utilize the information to predict the excited states.
An explicit example of this point is described in Sec.~\ref{sec:discussion} and \ref{apdx:non-linearity}.
Moreover, any $U_\mrm{ent}$ can be written in the form of a time-evolution operator as $e^{-iHT}$ under a certain Hamiltonian operator $H$.
In this formulation, one can naturally interpret the entangler as an operator evolving the single-qubit Pauli operators $\bigcup_{i=0}^{N-1}\{X_i, Y_i, Z_i\}$ into a linear combination of the multi-qubit ones in the Heisenberg picture, and the linear combination consists of more variety of the multi-qubit Paulis as the evolution takes a longer time (see the details in \ref{apdx:entangler-analysis}).
Second, the model is identical to quantum reservoir computing proposed in Ref.~\cite{Fujii2017} and quantum reservoir processing proposed in Ref.~\cite{Ghosh2019} if we choose the linear regression as the classical machine learning unit in the model.
One may also consider using general classical models such as the neural network, the Gaussian process regression, etc.
Even though the numerical simulations we carried out in this study leverage only a linear model, which actually gives sufficiently accurate predictions at least for the molecules we consider here, nonlinearity in the classical machine learning unit may be necessary to predict the excited states for certain tasks as discussed in Sec~\ref{sec:discussion} using an exactly-solvable toy model for the hydrogen molecule.
Third, as mentioned in the previous section, we stress that this scheme is very suitable to be combined with the VQE. 
The VQE finds a quantum circuit that produces an approximate ground state of a given Hamiltonian by using the variational principle and has been extended to obtain the excited states recently~\cite{McClean2017PRA, Colless2018, Nakanishi2019PRR, Parrish2019, Higgott2019, Jones2019, Ollitrault2019, Tilly2020}. 
Since it can handle Hamiltonians of large systems that are intractable by classical computers, the VQE is considered as one of the best approaches to utilize the NISQ devices for real-world problems.
In our quantum machine learning model, one can use the quantum circuit obtained by the VQE to make an input state (approximate ground state wavefunction) for the training and the prediction of our model.
There is no overhead cost at all to feed target data to the learning model in this case (see also Ref.~\cite{Uvarov2019}).
We review the VQE and one of its extensions to compute the excited states in~\ref{apdx:vqe}.

\subsection{Supervised learning of the model}
Next, we explain the procedure for supervised learning of our model.
First, 
we define the training set $\mathcal{R}$ whose elements $r\in \mathcal{R}$ are a set of characteristics of a molecule (e.g., name of a molecule and its atomic configuration), and we prepare the data  $\{\ket{\psi_0(r)}, \mathbf{y}(r)\}_{r\in \mathcal{R}}$ for training.
In the case of predicting excited states of a given Hamiltonian from its ground state, $\ket{\psi_0(r)}$ is the ground state of the molecular Hamiltonian $H(r)$ and $\mathbf{y}(r)$ contains the target properties of the excited states of $H(r)$, such as excitation energies.
Next, by using the training set, the classical machine learning unit $f_W$ is trained to predict $\{ \mathbf{y}(r) \}_{r\in \mathcal{R}}$ from the classical vectors $\{ \mathbf{x}_{\ket{\psi_0(r)}} \}_{r\in \mathcal{R}}$ which are calculated in the way described in the previous subsection.
A typical training algorithm for the supervised learning is to minimize a cost function defined to measure the deviations of the prediction $\{ f_W(\mathbf{x}_{\ket{\psi_0(r)}}) \}_{r\in \mathcal{R}}$ from the training data $\{ \mathbf{y}(r) \}_{r\in \mathcal{R}}$ by tuning $W$.

We note that our model is easier to be trained and less costly in terms of the number of runs of quantum computers compared with the so-called ``quantum circuit learning" where parameters of the quantum circuit are optimized~\cite{Mitarai2018, Farhi2018, Havlicek2019, Kusumoto2019} since once the classical representation of the quantum state $\{ \mathbf{x}_{\ket{\psi_0(r)}} \}_{r\in \mathcal{R}}$ is obtained, there is no need to run the quantum device afterwards for training the model.

\section{Numerical demonstration for small molecules \label{sec:result}}
In this section, we numerically demonstrate the ability of our model to reproduce excited-state properties from the ground state wavefunctions by taking small molecules as examples.
We consider three types of molecules: LiH molecule, $\mrm{H_4}$ molecule whose hydrogen atoms are aligned linearly with equal spacing, and $\mrm{H_4}$ molecules whose hydrogen atoms are placed in a rectangle shape.
We call them as LiH, $\mrm{H_4}$ (line), $\mrm{H_4}$ (rectangle), respectively.
We evaluate our model in two situations, in one of which ideal outputs of the quantum circuits are available (\textit{noiseless}), and inevitable noise in the real NISQ devices is considered in the other (\textit{noisy}).
The electronic ground states of those molecules with various atomic geometries are prepared by diagonalizing the Hamiltonian for the noiseless simulation and by numerically simulating the VQE for the noisy simulation.
Then, we train our model with the linear regression as its classical machine learning unit to predict the first and second excitation energies and the transition dipole moment among them whose values are obtained by exactly solving the Hamiltonian.
Numerical results show that our model can properly reproduce the excited states and illustrate the predictive power of our model.

\subsection{Dataset\label{sec:dataset}}
To prepare a dataset for the simulations, we consider the electronic Hamiltonians of the following configurations.
For LiH molecule and $\mrm{H_4}$ (line), the atomic distances are in the range of $[0.5\mrm{\AA}, 3.3\mrm{\AA}]$.
For $\mrm{H_4}$ (rectangle), we choose the two spacing of atoms (lengths of two edges) in $[0.5\mrm{\AA}, 2.0\mrm{\AA}]\times[0.5\mrm{\AA}, 2.0\mrm{\AA}]$.
We perform the standard Hartree-Fock calculation by employing the STO-3G minimal basis and construct the fermionic second-quantized Hamiltonian for all of the molecules and configurations~\cite{McArdle2018, Cao2019} with open-source libraries PySCF~\cite{PySCF} and OpenFermion~\cite{McClean2020}.
Two Hartree-Fock orbitals with the highest and the second-highest energies among six orbitals of LiH molecule are removed by assuming they are vacant because they are composed almost completely from 2$p_x$ and 2$p_y$ atomic orbitals of LiH and do not significantly contribute to the binding energy of LiH.
Then the Hamiltonian is mapped to the sum of the Pauli operators by the Jordan-Wigner transformation~\cite{Jordan1928} which we denote $H(r)$
(a review of the Jordan-Wigner transformation is given in~\ref{apdx:jw-transformation}).
Then, the electric Hamiltonians for all of the molecules turn into 8-qubit Hamiltonians.  

The training and test datasets for the simulations are prepared for each Hamiltonian $H(r)$ in the following way.
First, in the case of the noiseless simulation, the ground state of $H(r)$ is prepared by the exact diagonalization.
In the case of the noisy simulation, the VQE algorithm is applied to $H(r)$, and the approximate ground state is obtained as $\ket{\tilde{\psi}_0(r)} = U(\vec{\theta})\ket{0}$. 
Here $U(\vec{\theta})$ is a variational quantum circuit (ansatz) with classical parameters $\vec{\theta}$ and $\ket{0}$ is a reference state.
We adapt the unitary coupled-cluster singles and doubles ansatz~\cite{Peruzzo2014, Lee2019} as $U(\vec{\theta})$.
Next, we compute the quantities of the excited-state properties to be predicted,
\begin{equation}
 \mathbf{y}(r) = ( \Delta E_1(r), \Delta E_2(r), \|\bm{\mu}_\mrm{eg}(r)\| )^T,
\end{equation}
where $\Delta E_{1(2)}(r) = E_{1(2)}(r) - E_0(r)$ is the first (second) excitation energy of $H(r)$ in the sector of neutral charge, where $E_{0,1,2}(r)$ are three lowest eigenenergies of $H(r)$ in the same sector ignoring degeneracy.
For our choice of the molecules and configurations, $E_0(r)$ is the energy of the spin-singlet ground state $S_0$, and $E_1(r)$ is the energy of the spin-triplet excited state $T_1$.
$E_2(r)$ is the energy of the spin-singlet excited state $S_1$ or the spin-triplet excited state $T_2$ depending on the configurations of the molecule.
The transition dipole moment between the ground state and the excited state $\bm{\mu}_\mrm{eg}(r)$ is defined as
\begin{equation}
 \bm{\mu}_\mrm{eg}(r) =  \braket{\psi_0(r) |\bm{\mu}| \psi_\mrm{ex}(r)},
\end{equation}
where $\ket{\psi_0(r)}$ is the exact ground state of $H(r)$ (the singlet state $S_0$), $\ket{\psi_\mrm{ex}(r)}$ is the exact excited state of $H(r)$ which has the lowest energy among those having a non-zero transition dipole moment from the ground state (typically $S_1$ state), and $\bm{\mu} = -e (\hat{x},\hat{y},\hat{z})^T $ is the dipole moment operator with electronic charge $e$. In this study, we use its L2-norm $\|\bm{\mu}_\mrm{eg}(r)\|$ for the learning tasks. 
The calculation of each value of $\mathbf{y}(r)$ is performed by the exact diagonalization of $H(r)$ for both of the noiseless and noisy simulations.
To stabilize the learning process, we scale those calculated values to fit them into the $[-1, 1]$ range, so that the maximum value $y^{(k)}_{\max}$ and the minimum value $y^{(k)}_{\min}$ in the training dataset are scaled as $y^{(k)}_{\max} = \max_{r\in \mathcal{R}}y^{(k)}(r) \rightarrow 1$ and  $y^{(k)}_{\min} = \min_{r\in \mathcal{R}}y^{(k)}(r) \rightarrow -1$ where $y(r)^{(k)}$ denotes the $k$-th element of $\bm{y}(r)$ for each $k = 1, 2, 3$, and other values, including those in the test dataset, are mapped as $y(r)^{(k)} \rightarrow 2\frac{y(r)^{(k)}-y^{(k)}_{\max}}{y(r)^{(k)}_{\max} - y(r)^{(k)}_{\min}}-1$.

For the numerical experiments,
we randomly split those obtained data $\{\ket{\tilde{\psi}_0(r)}, \mathbf{y}(r) \}_{r}$ into the training set and the test set for the evaluation of the model. 
We used 30 training data points and 50 test points, respectively for the tasks of the LiH and $\mathrm{H_4}$ (linear) molecules, and 250 training data points and 1250 test data points for the $\mathrm{H_4}$ (rectangle) molecules.

\subsection{Model for the simulations}
The entangler $U_\mrm{ent}$ in the model is chosen to be the time-evolution operator $e^{-iH_\mrm{TFIM}T}$ under the random transverse-field Ising model (TFIM),
\begin{equation}
 H_\mrm{TFIM} = \sum_{i,j=0}^{N-1} J_{ij} Z_i Z_j + \sum_{i=0}^{N-1} h_i X_i, 
\end{equation}
where $X_i$ and $Z_j$ are Pauli operators acting on the site $i, j$-th qubit, coefficients $h_i$ and $J_{ij}$ are sampled from the Gaussian distributions $N(1, 0.1)$ and $N(0.75, 0.1)$, respectively, and we set $T=10$.
These coefficients are fixed during each of the numerical simulations.
This type of the entangler can be implemented on various types of the NISQ devices; for example, in the case of superconducting qubits, it can be realized by a sequence of the cross resonance gates~\cite{Kandala2017,Kandala2019} or simply tuning the resonance frequency of the qubits~\cite{Havlicek2019}.
We note that a similar kind of the quantum reservoir has recently been implemented on a real NISQ device~\cite{Chen2020temporal}. 

In our numerical simulations, this time evolution is exactly simulated as the unitary operation $e^{-iH_\mrm{TFIM}T}$ acting on the input state. 

For the classical machine learning unit for the numerical demonstration, we employ the linear regression (LR)~\cite{Bishop2006}. 
Although the LR does not have nonlinearity which is in principle necessary to compute the excited-state properties (see Sec.~\ref{sec:discussion}), it performs well enough for the molecular Hamiltonians we consider for the simulations as shown in Sec. \ref{subsec:demo_result}, so it serves as a nice demonstrative model to evaluate the concept of our model. 

The output function of the LR is
\begin{equation}
 f^{(k)}(\mathbf{x}_{\ket{\psi}}) = \mathbf{w}_\mrm{out}^{(k)} \cdot \mathbf{x}_{\ket{\psi}},
\end{equation}
where $\mathbf{w}_\mrm{out}^{(k)}$ is a $3N$-dimensional vector, or parameters of the model, to be optimized, and $k=0,1,2$ corresponds to the component of the prediction for $\mathbf{y}=(y^{(0)}, y^{(1)}, y^{(2)})^T$. 
The model is trained to minimize the mean squared error (MSE) cost function
\begin{equation} \label{cost_LR}
    L_\mrm{LR}(\{ \mathbf{w}_\mrm{out}^{(k)}\} )
    = \frac{1}{|\mathcal{R}|}\sum_{r \in \mathcal{R}}
     \left| \mathbf{w}_\mrm{out}^{(k)} \cdot \mathbf{x}_{\ket{\psi_0(r)}} - y(r)^{(k)} \right|^2,
\end{equation}
where $\mathcal{R}$ represents the training dataset, respectively for each target property.
The exact optimum of the cost function can be obtained as
\begin{equation} 
    \textbf{w}^{(k)*}_\mrm{out} = \left( V^T V \right)^{-1} V^T \mathbf{Y}^{(k)},
\end{equation}
where $V$ is a $|\mathcal{R}|\times 3N$ dimensional matrix whose $i$-th row is ${ \mathbf{x}_{\ket{\psi_0(r_i)}}}^T$, and $\mathbf{Y}^{(k)}$ is a $|\mathcal{R}|$ dimensional column vector whose $i$-th component is $y(r_i)^{(k)}$, where $r_i$ is the $i$-th element of $\mathcal{R}$.
The whole classical process requires the computational complexity of $\mathcal{O}(N^3+N^2|\mathcal{R}|)$.

\subsection{Simulation of quantum circuits \label{subsec:simulation_def}}
To check the practical advantage of our model with the NISQ devices, we numerically simulate quantum circuits of the model considering the noiseless and noisy situations (including preparation of the ground state wavefunctions by the VQE for the noisy case).
The latter reflects a more realistic situation of experiments on a real NISQ device, but we stress that the former still serves as a reference point to judge whether the model has the capability of performing the learning task or not.

In the noiseless simulation, the expectation value of the Pauli operator $\braket{\psi|P_i|\psi}$, where the $\ket{\psi}$ is a quantum state and $P_i$ is the Pauli operator acting on $i$-th qubit, is estimated exactly by calculating the inner product.
In the noisy simulation, we consider two error sources which make estimations of those expectation values deviate from the exact ones.
One of them is a sequence of the depolarizing noise channels~\cite{Nielsen2011} that transform the quantum state $\rho = U_\mrm{ent}\ket{\psi}\bra{\psi}U_\mrm{ent}^\dag$ from the reservoir into $\rho' = \mathcal{E}_{N-1}\circ \ldots\circ\mathcal{E}_0(\rho)$, where $\mathcal{E}_i$ is the depolarizing channel that acts on $i$-th qubit as $\mathcal{E}_i(\sigma) = (1-p)\sigma + \frac{p}{3} (X_i\sigma X_i + Y_i\sigma Y_i + Z_i\sigma Z_i)$.
We take $p=0.01$ in the simulations.
The other source is the so-called shot noise that stems from the finite number of shots for the projective measurements of the Pauli operator $P_i$.
Each measurement returns $\pm 1$ according to the probability distribution determined by the exact values of $\mrm{Tr}\left(\rho' P_i\right)$.
We sample $10^4$ shots of measurements to compute the ground state with the VQE, and $10^6$ shots for each Pauli operator to construct the vector in Equation~\eqref{eq:x-vec}. These are feasible numbers in experiments~\cite{Kandala2017}.

\subsection{Results \label{subsec:demo_result}}
The model described in the previous subsections is trained by the training dataset and evaluated by the test set.
The evaluation is performed based on the mean absolute error (MAE) for the test set $\mathcal{T}$,
\begin{equation} \label{loss_test}
    C_\mathcal{T} =
    \frac{1}{3|\mathcal{T}|}\sum_{r \in \mathcal{T}}\sum_{k=0}^2
    \big| \mathbf{f}_W(\mathbf{x}_{\ket{\psi_0(r)}}) - \mathbf{y}(r)^{(k)} \big|,
\end{equation}
where $\mathbf{f}_W(\mathbf{x})$ is the output of the model considered as a vector.
We train and evaluate the model for each molecule separately.

\subsubsection{Noiseless simulation\label{subsubsec:noiseless_result}}
\begin{figure*}
    \begin{center}
    \includegraphics[width=.45\textwidth]{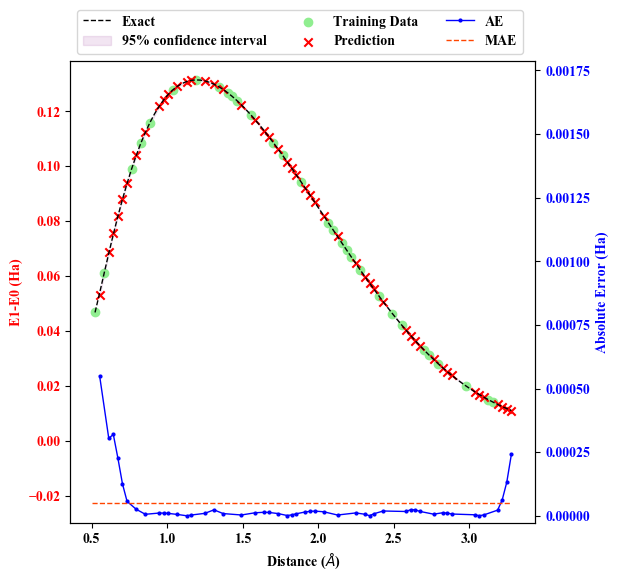} 
    \includegraphics[width=.45\textwidth]{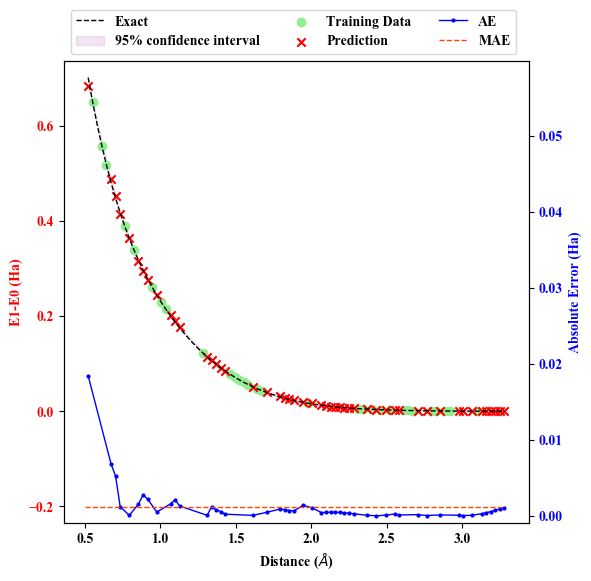} \\
    \includegraphics[width=.45\textwidth]{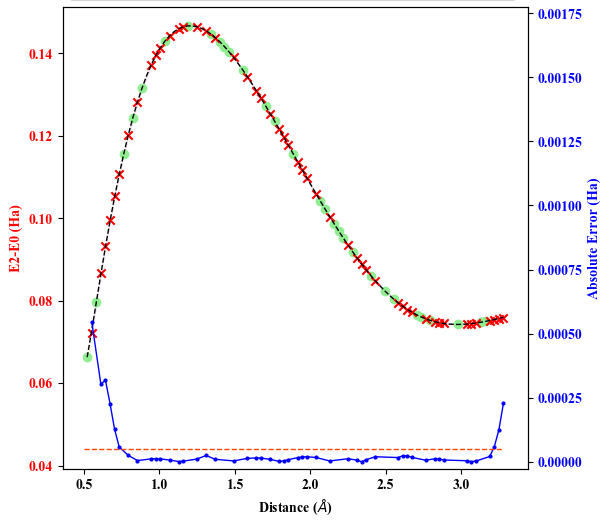} 
    \includegraphics[width=.45\textwidth]{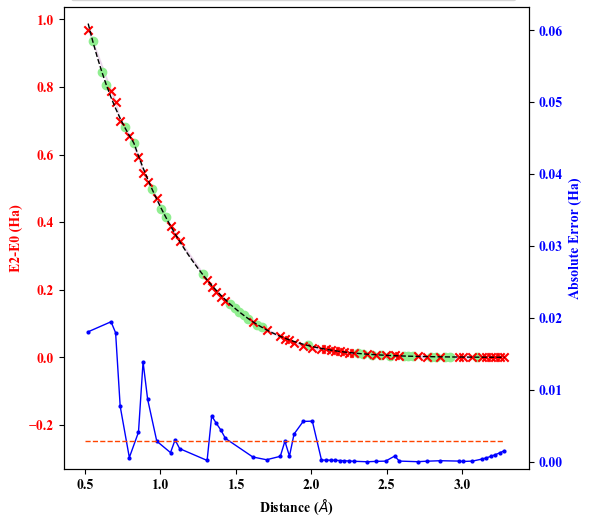} \\ 
    \includegraphics[width=.45\textwidth]{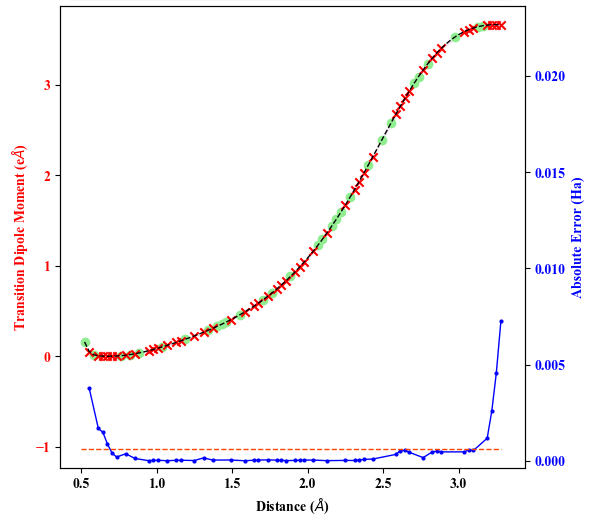}  
    \includegraphics[width=.45\textwidth]{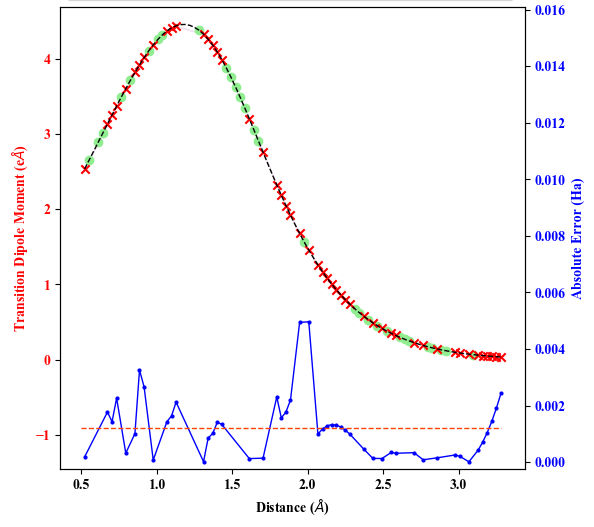}
    \end{center}
     \caption{\label{fig:LR_LiH_H4line_exact}
     The  prediction  results  by  the  trained  model for LiH (left column) and $\mrm{H_4}$ (line) (right column) for the noiseless simulations. Top, middle, and bottom panels display the first, second excitation energies and the transition dipole moment, respectively.
     The green circles represent the training data points and the red crosses are the predictions. The exact values are displayed as the black line. Those values are read from the left ticks of each panel. The blue plot lines with circles represent the absolute errors between the predictions and the exact values, and the orange line indicates their mean. These error values are read from the right ticks of each panel.}
\end{figure*}

\begin{figure*}
\begin{center}
     \includegraphics[width=.95\textwidth]{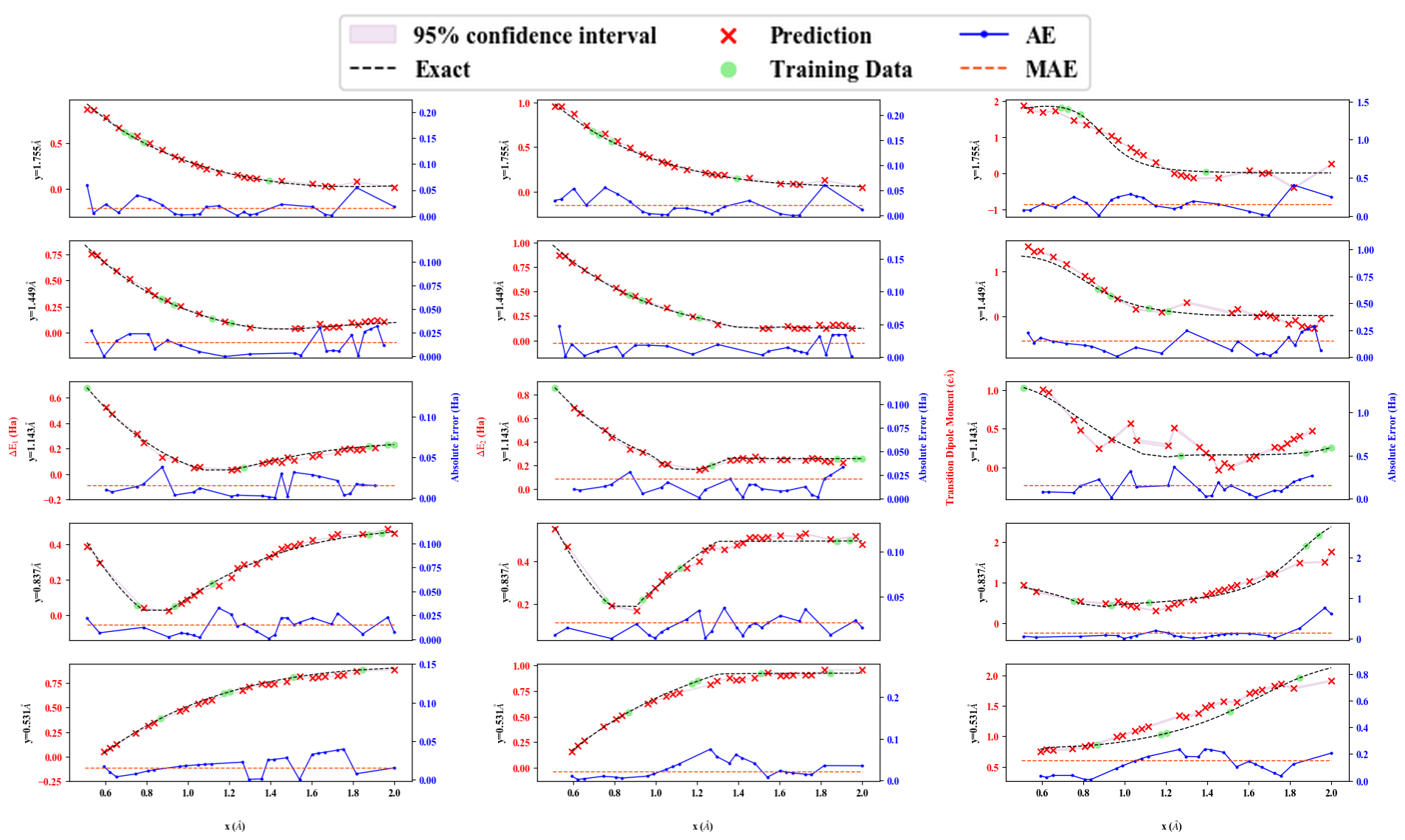}
\end{center}
  \caption{\label{fig:LR_H4rect_exact} The same figures as Figure~\ref{fig:LR_LiH_H4line_exact} for $\mrm{H_4}$ (rectangle).
  $x$ and $y$ denote the 
  two atomic spacings of the rectangular geometry.}
\end{figure*}

The prediction results by the trained model in the noiseless numerical simulation are shown in Figures~\ref{fig:LR_LiH_H4line_exact} and~\ref{fig:LR_H4rect_exact}.
In these figures, the excited-state properties $\mathbf{y}(r)$ are scaled back to the original scale.
Our model obviously reproduces the exact values of the excited-state properties $\mathbf{y}(r)$ for all of the three molecule types.

To quantify it, in the upper rows of Table~\ref{Tab:chem-accuracy}, we summarize the MAEs of $\Delta E_1$ and $\Delta E_2$ for the test data, Eq.~(\ref{loss_test}),  between the predictions and the exact values scaled back to the original scale.
For the LiH and $\mrm{H_4}$ (linear) molecules, the MAEs are below or in the comparable scale to the chemical accuracy $1.6\times 10^{-3} \mrm{Ha}$. 
The error is larger than the chemical accuracy for $\mrm{H_4}$ (rectangle), and it is probably because the degrees of freedom of the molecular structure of $\mrm{H_4}$ (rectangle) are larger, and the LR model may not have a sufficient expressive space. 
Utilizing other machine learning methods (e.g. neural networks) is one possible way to achieve more accurate results.

Also, we investigate the necessity of the entangler $U_\mrm{ent}$ by comparing the values of the MAE (Equation~\eqref{loss_test}) for the test set after the training.
The evaluations of the learners with and without the entangler are summarized in Table~\ref{Tab:LR}, indicating that the entangler significantly enhances the predictive power of the model.
To treat all the excited-state properties on an equal footing, here we use the scaled values of $\mathbf{y}(r)$.
In Sec.~\ref{sec:discussion}
and~\ref{apdx:non-linearity}, another supporting result for the necessity of the entangler is presented by using an exactly solvable model for the hydrogen molecule.

These results from the noiseless simulations illustrate the predictive power of our model for the difficult task to predict the excited-state properties only from the ground state.

\begin{table*}
\caption{ \label{Tab:chem-accuracy}
The individual MAEs estimating the properties $\Delta E_1$ and $\Delta E_2$ in the units of Hartree, compared with the chemical accuracy $1.6 \mrm{mHa}$ for both of the noisy and the noiseless cases. }
\begin{indented}
\item[]\begin{tabular}{r
r
>{\raggedleft\arraybackslash}p{0.11\linewidth} 
>{\raggedleft\arraybackslash}p{0.11\linewidth} 
>{\raggedleft\arraybackslash}p{0.18\linewidth}
>{\raggedleft\arraybackslash}p{0.15\linewidth}}
\toprule
     & & LiH      & H4 (line) & H4 (rectangle) & Chemical Accuracy\\ 
\hline
\multirow{2}{*}{MAE (noiseless) (mHa)}&
    $\Delta E_1$ & 0.1 & 1.4 & 21.3     & \multirow{4}{*}{1.6} \\ 
    &$\Delta E_2$ & 0.1 & 3.2 & 28.0     & \\ 
\multirow{2}{*}{MAE (noisy) (mHa)}&
    $\Delta E_1$ & 15.8 & 39.2 & 109.1     &  \\
    &$\Delta E_2$ & 15.4 & 50.3 & 103.8     &  \\ 
\bottomrule
\end{tabular}
\end{indented}
\end{table*}

\begin{table*}
\caption{ \label{Tab:LR}
The MAEs evaluated with the test set for the trained models in the noiseless situation with and without the entangler. The mean values are taken over all the three properties to be estimated.
The output values and the excited-state properties are in the same standardized scale as explained in Section~\ref{sec:dataset}.
The MAEs for the random guess are also presented as a reference.
}
\begin{indented}
\item[]\begin{tabular}{r 
p{0.15\linewidth} 
p{0.15\linewidth} 
p{0.18\linewidth}}
\toprule
     & LiH      & H4 (line) & H4 (rectangle) \\ 
\hline
Test MAE with entangler $U_\mrm{ent}$       & 0.0181       & 0.0203        & 0.0836 \\ 
Test MAE without entangler $U_\mrm{ent}$    &  0.172      &  0.324           & 0.300  \\ 
Random Guess      & 0.673    & 0.544       & 0.444 \\
\bottomrule
\end{tabular}
\end{indented}
\end{table*}

\subsubsection{Noisy simulation\label{subsubsec:noisy_result}}
In order to evaluate our scheme in a realistic situation with a quantum device, we add two noise sources to the simulation as described in Sec.~\ref{subsec:simulation_def}. 
In this case, to enhance the noise robustness, we make two modifications to the noiseless case as follows. 
First, after obtaining the classical vectors $\{\mathbf{x}_{\ket{\psi_0(r)}}\}_{r\in\mathcal{R}}$ by processing the training dataset with the noisy quantum circuit, we make 100 copies of every vector $\mathbf{x}_{\ket{\psi_0(r)}}$ on a classical computer, and add a gaussian noise $N(0, 2\times 10^{-3})$ to each component of it. We stack these vectors, and now we have a new $100|\mathcal{R}|\times 3N$
dimensional matrix $V'$. The vector $\mathbf{Y}^{(k)}$ is also duplicated 100 times to match up with $V'$ (let us call this new vector $\mathbf{Y}'^{(k)}$ for later use). Notice that this modification does not affect the required number of measurements of the quantum circuit. 
Second, we add the L2 regularization term into the cost function of the LR, particularly saying the cost function becomes
\begin{align} \label{regulatization}
     L'_\mrm{LR}(\{ \mathbf{w}_\mrm{out}^{(k)}\} )
    = L_\mrm{LR}(\{ \mathbf{w}_\mrm{out}^{(k)}\} )
    + \alpha \|\mathbf{w}_\mrm{out}^{(k)}\|^2,
\end{align}
where we used $\alpha = 10^{-3}$ in the simulations. We may obtain the exact optimum by computing
\begin{equation} 
    \textbf{w}^{(k)*}_\mrm{out} = \left( V'^T V'  + \alpha I \right)^{-1} V'^T \mathbf{Y}'^{(k)},
\end{equation}
where $I$ is an identity matrix of $3N\times 3N$ dimensions. Both of these two modifications work as regularizations preventing the model from overfitting due to the outliers with large noises. 

\begin{figure*}
\begin{center}
    \includegraphics[width=.45\textwidth]{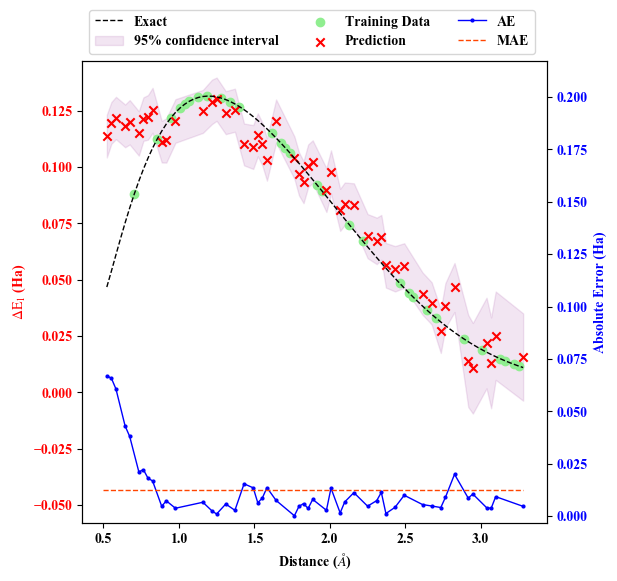} 
    \includegraphics[width=.45\textwidth]{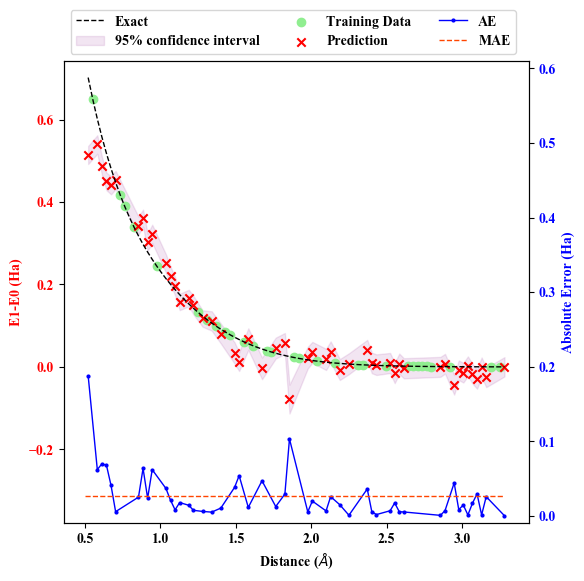} \\
    \includegraphics[width=.45\textwidth]{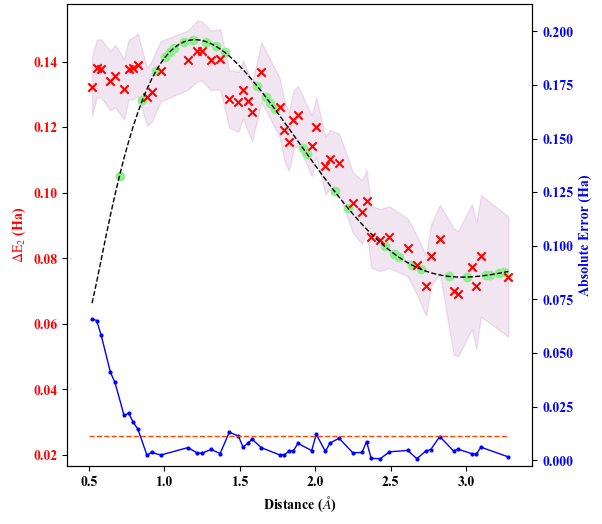} 
    \includegraphics[width=.45\textwidth]{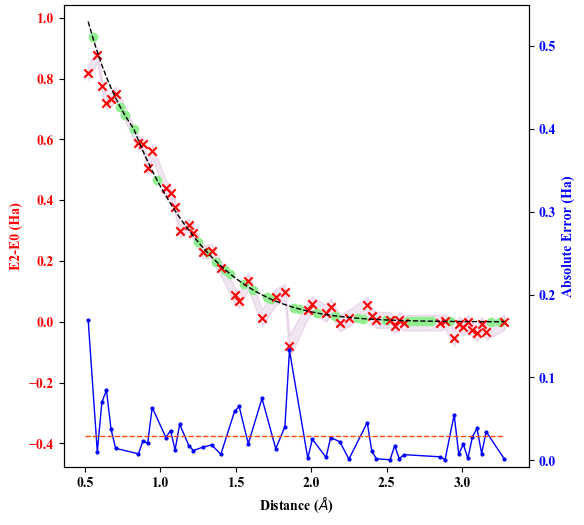} \\ 
    \includegraphics[width=.45\textwidth]{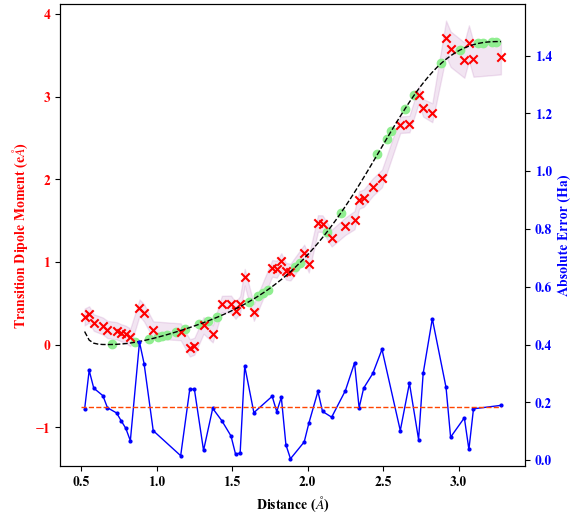}  
    \includegraphics[width=.45\textwidth]{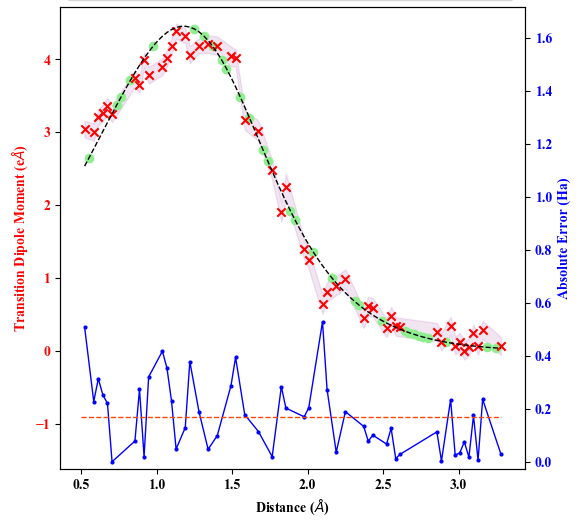}
    \end{center}
     \caption{\label{fig:LR_LiH_H4line_noisy}
     The  prediction  results  by  the  trained  model for LiH (left column) and $\mrm{H_4}$ (line) (right column) for the noisy simulations. Top, middle, and bottom panels display the first, second excitation energies and the transition dipole moment, respectively. The green circles represent the training data points and the red crosses are the predictions. The exact values are displayed as the black line. Those values are read from the left ticks of each panel. The blue plot lines with circles represent the absolute errors between the predictions and the exact values, and the orange line indicates their mean. These error values are read from the right ticks of each panel.}
\end{figure*}

\begin{figure*}
    \includegraphics[width=.95\textwidth]{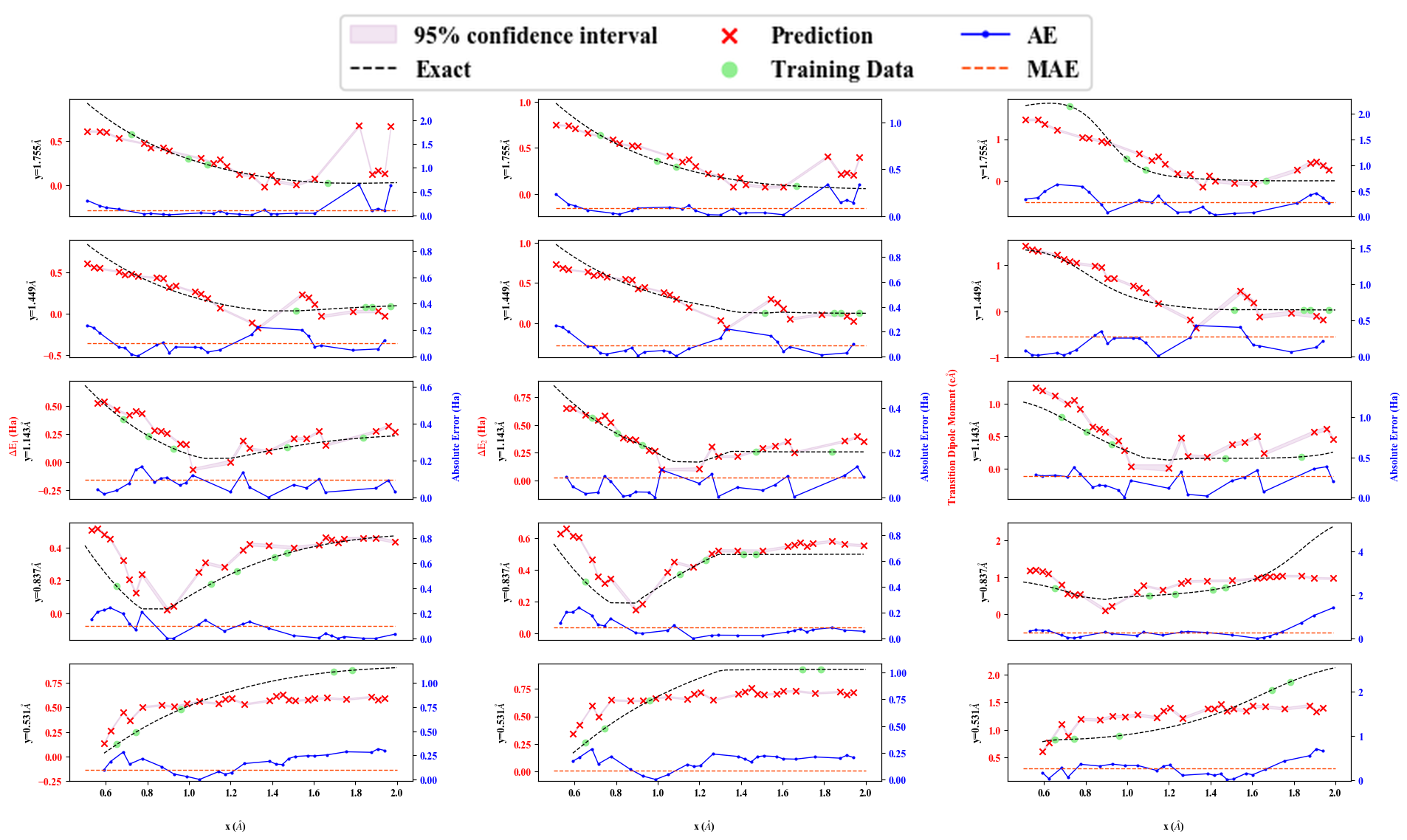}
  \caption{\label{fig:LR_H4rect_noisy} The same figures as Figure ~\ref{fig:LR_LiH_H4line_noisy} for $\mrm{H_4}$ (rectangle).
  $x$ and $y$ denote two atomic spacings of the rectangular geometry.}
\end{figure*}

The prediction results are presented in Figures~\ref{fig:LR_LiH_H4line_noisy} and \ref{fig:LR_H4rect_noisy}.
We see that the model still predicts $\mathbf{y}(r)$ well even in this noisy case.

We attribute this noise-robustness to the regularization technique of the LR in Equation~(\ref{regulatization}).
The MAEs for the predictions of $\Delta E_1$ and $\Delta E_2$ are summarized in the lower rows of Table~\ref{Tab:chem-accuracy} in the same way as the noiseless cases. 
The noise makes the accuracy of the predictions worse than those of the noiseless cases, and
all of the errors become larger than the chemical accuracy.
A part of the reason for this is because the noise hinders obtaining sufficiently precise ground states of the molecular Hamiltonians. Indeed, for example in the case of LiH, we find that the ground-state energies computed by the VQE in the noisy situation already have a larger error (0.0074 Ha) than the chemical accuracy, and the MAEs of the predictions ($\sim 0.015$ Ha) are in the similar order.
Adapting the error mitigation techniques~\cite{Temme2017, Endo2018} to the VQE can remove the effect of the noise and will yield more accurate results even in the noisy situation.
In~\ref{apdx:vqe-shots}, we present how the accuracy of the predictions for the excited-state properties varies as the function of the number of the shots used to perform the VQE.

\section{Discussion \label{sec:discussion}}
\subsection{Necessity of the entangler $U_\mrm{ent}$ and the nonlinearity of $f_W$\label{subsec:hubbard}}
Here we discuss the necessity of the entangler $U_\mrm{ent}$ and the nonlinearity in the classical machine learning unit $f_W$ by considering an exactly solvable model of fermions, namely, the 2-site fermion Hubbard model at half-filling~\cite{Hubbard1963}.

The 2-site Hubbard model is defined as
\begin{align} \label{eq:Hubbard}
 H_\mathrm{Hub}(U) = - &\sum_{\sigma=\uparrow,\downarrow} \left( c_{0,\sigma}^\dagger c_{1,\sigma} + \rm{h.c.} \right)+ U \sum_{i=0,1}  n_{i,\uparrow} n_{i,\downarrow} 
 \end{align}
where $c_{i, \sigma}, c_{i, \sigma}^\dagger$ are fermionic creation and annihilation operators acting on an electron with spin $\sigma = \uparrow, \downarrow$ located at $i$-th site $(i=0,1)$, and $ n_{i,\sigma} = c_{i, \sigma}^\dagger c_{i, \sigma}$ is the number operator of an electron with spin $\sigma$ at $i$-th site.
The parameter $U>0$ determines the strength of electron repulsion.
This system can be considered as a simplified model of a hydrogen molecule whereas it also serves as a prototype of strongly-correlated materials.
When we restrict ourselves into the sector where the number of electrons is two, i.e. the neutral hydrogen states, the first, second excitation energies and the transition dipole moment are
\begin{align}
 \Delta E_1 &= \frac{U}{2} \left(- 1 + \sqrt{1+\frac{16}{U^2}} \right),  \\
 \Delta E_2 &= \frac{U}{2} \left( 1 + \sqrt{1+\frac{16}{U^2}} \right), \\
 \|\bm{\mu}_\mrm{eg}\| &= \sqrt{\frac{1-(1+16/U^2)^{-1/2}}{2}}, 
\end{align}
respectively.
We note that the dipole moment operator is defined as $\bm{\mu} = \frac{1}{2}(n_{1,\uparrow}+n_{1,\downarrow}-n_{0,\uparrow}-n_{0,\downarrow})$.

Applying the Jordan-Wigner transformation~\cite{Jordan1928} to the system (Equation~\eqref{eq:Hubbard}), we obtain the 4-qubit Hamiltonian $H_\mrm{qubit}(U)$.
We denote the ground state of $H_\mrm{qubit}(U)$ as $\ket{\psi_0(U)}$.
When there is no entangler, the classical vector $\mathbf{x}_{\ket{\psi_0(U)}}$ is trivial because
\begin{equation}
\braket{X_j}_\mrm{GS} = \braket{Y_j}_\mrm{GS} = \braket{Z_j}_\mrm{GS} = 0
\end{equation}
holds for all qubit sites $j=0,1,2,3$, where we define $\braket{\ldots}_\mrm{GS}= \braket{\psi_0(U)|\ldots|\psi_0(U)}$.
In contrast, when there is an entangler $U_\mrm{ent}$ in our model, it converts the Pauli operators $X_j, Y_j, Z_j$ into a sum of more complicated Pauli strings as $U_\mrm{ent}^\dag Z_0 U_\mrm{ent} = Z_0Z_1 + 0.2 Z_1 X_1 Z_2 Y_2 + \ldots$ in the Heisenberg picture.
Then, the classical vector $\mathbf{x}_{\ket{\psi_0(U)}}$ contains contributions from the terms like $\braket{Z_0Z_1}_\mrm{GS}$.
It follows that
\begin{equation}
  \braket{Z_0Z_1}_\mrm{GS} = - \left(1+\frac{16}{U^2}\right)^{-1/2},
\end{equation}
and
\begin{align}
 \Delta E_1  &= \frac{2\left(1+\braket{Z_0 Z_1}_\mrm{GS}\right)}{ \sqrt{1 - \braket{Z_0 Z_1}_\mrm{GS}^2} },  \\
\Delta E_2 &= \frac{2\left(1-\braket{Z_0 Z_1}_\mrm{GS}\right)}{ \sqrt{1 - \braket{Z_0 Z_1}_\mrm{GS}^2} },
\\
 \|\bm{\mu}_\mrm{eg}\| &= \sqrt{\frac{1+\braket{Z_0 Z_1}_\mrm{GS}}{2}}.
\end{align}
These equation indicate that the excitation energies can be predicted by utilizing the values of $\braket{Z_0Z_1}_\mrm{GS}$ appropriately.
Therefore, one can see that it is possible to predict the excitation energies from the classical vector $\mathbf{x}_{\ket{\psi_0(U)}}$ if the ground state vector is processed by an entangler, and the classical machine learning unit $f_W$ has enough nonlinearity.
These equations also imply that the details of the entangler, which determine coefficients of the terms like $\braket{Z_0Z_1}_{\mrm{GS}}$ in $\mathbf{x}_{\ket{\psi_0(U)}}$, is not so important for predictions; the classical machine learning unit can compensate the difference of such coefficients.
In Appendix, we provide further analysis of the 2-site Hubbard model including the necessity of nonlinearity.

\subsection{Generalizablity}
In the numerical simulations in Sec.~\ref{sec:result}, the models are trained and evaluated for each molecule separately.
The generalizability of our model to predict the properties of various molecules simultaneously is one possibility of our model for future extensions.

To make our model more powerful and capable of taking various molecules as inputs, several modifications can be considered.
First, including the ground state energy which can also be calculated by the VQE besides the ground state in the input of the classical machine learning unit $f_W$ will be necessary since otherwise, one may not determine the energy scale of an input molecule.
Second, replacing the entangler $U_\mrm{ent}$ with a parametrized quantum circuit $V(\theta)$ and optimizing the circuit parameters $\theta$ along with the classical machine learning unit increase the degree of freedoms of the model and may result in a better predictive power, 
with a possible drawback that the number of required experiments on the NISQ devices would increase in the training step. 
Exploring these ideas is an interesting future direction of the work.

\section{Conclusion \label{sec:conclusion}}
In this study, we introduce a new quantum machine learning framework for predicting the excited-state properties of a molecule from its ground state wavefunction.
By employing the quantum reservoir and choosing simple one-qubit observables for measurements accompanied by post-processing with classical machine learning,
one may process 
our framework easily on the NISQ devices requiring the realistic number of runs of them.
The numerical simulations with and without the noise in outputs of quantum circuits demonstrate that our model accurately predicts the excited states.
Although our framework is tested only with small molecules to illustrate its potential in the numerical simulations,
we expect that it will benefit the calculation of excited states of larger molecules by reducing the computational cost from calculating exact solutions. 
Our result opens up the further possibility to utilize the NISQ devices in the study of quantum chemistry and quantum material fields.

\section*{Acknowledgement}
HK was supported by QunaSys Inc.
HK and YON acknowledge Suguru Endo, Kosuke Mitarai, Nobuyuki Yoshioka, Wataru Mizukami, and Keisuke Fujii for valuable discussions.
This work was also supported by MEXT Q-LEAP JPMXS0118068682.

\appendix
  
\section{Variational Quantum Eigensolver (VQE) and its extension to the excited states \label{apdx:vqe}}
In this Appendix, we first review the VQE algorithm~\cite{Peruzzo2014} which finds the ground state of a given Hamiltonian by using the near-term quantum computers.
We use it to prepare the ground states of the molecular Hamiltonians considering the realistic noisy situation in Sec.~\ref{sec:result}.
Next, to give the readers an insight on how costly it is to find the excited states of a given Hamiltonian on the near-term quantum computers compared with the computations for the ground states, we review the subspace-search VQE (SSVQE) algorithm~\cite{Nakanishi2019PRR} as one example of such algorithms.

\subsection{VQE algorithm}
The VQE tries to compute the minimum eigenvalue and its corresponding eigenstate of a given observable $H$ by minimizing the expectation value of $H$ with the ansatz state $\ket{\psi(\theta)}=U(\vec{\theta})\ket{0}$, where $U(\vec{\theta})$ is the parameterized unitary circuit on a quantum computer with classical parameters $\vec{\theta}$ and $\ket{0}$ is some reference state.
When the expectation value $E(\vec{\theta}) = \braket{\psi(\vec{\theta})|H|\psi(\vec{\theta})}$ reaches the minimum at $\vec{\theta}_\mrm{opt}$ by optimizing the parameters $\vec{\theta}$, $E(\vec{\theta}_\mrm{opt})$ and $\ket{\psi(\vec{\theta}_\mrm{opt})}$ are the closest approximation of the lowest eigenvalue and its corresponding eigenstate, respectively.
Evaluation of $E(\vec{\theta})$ for a given $\vec{\theta}$ is performed by the near-term quantum computers, and one uses a classical optimization algorithm to iteratively update the values of $\vec{\theta}$ to find the minimum.
This classical-quantum hybrid architecture of the VQE algorithm requires less computational/experimental abilities for quantum computers than the long-term, pure-quantum algorithms such as the phase estimation, so that one may run it on the near-term quantum computers.

When applying the VQE to the molecular Hamiltonian, first we prepare the observable $H$ as the second-quantized Hamiltonian of a given molecule by using the finite number of orbitals. Typically, the Hartree-Fock molecular orbitals are used for the second-quantization and each spin orbital corresponds one qubit~\cite{McArdle2018, Cao2019}.
Since the second-quantized Hamiltonian is written in fermionic operators while quantum computers can handle with qubit operators $P \in \{I, X, Y, Z\}^{\otimes N}$ only, it is then mapped into the linear combination of qubit operators, $H \rightarrow \sum_P h_P P$ where $h_P$ is a coefficient corresponding to the operator $P$.
One example of such the fermion-spin mapping is the Jordan-Wigner transformation which is reviewed in~\ref{apdx:jw-transformation}.

\subsection{Subspace-search variational quantum eigensolver (SSVQE) for excited states}
Here, we also review the SSVQE algorithm~\cite{Nakanishi2019PRR}, which is one of the algorithms to find the eigenstates corresponding to the higher eigenvalues of an observable $H$ on the near-term quantum computers.
Suppose we would like to find the $k$ lowest eigenvalues and eigenstates of $H$.
The SSVQE employs the $k$ reference states  $\{\ket{\phi_i}\}_{i=1}^k$ which are mutually orthogonal and prepares the $k$ ansatz states with a parameterized unitary circuit $U(\vec{\theta})$ as $\{\ket{\psi_i(\vec{\theta})} = U(\vec{\theta})\ket{\phi_i} \}_{i=1}^k$. 
It was shown in~\cite{Nakanishi2019PRR} that when the following cost function
\begin{equation}
    C(\vec{\theta}) = \sum_{i=1}^k w_i \braket{\psi_i(\vec{\theta})|H|\psi_i(\vec{\theta})}
\end{equation}
takes the minimum at $\vec{\theta}_\mrm{opt}$ for appropriate weights $\{w_i\}_{i=1}^k$ satisfying $i < j \Rightarrow w_i > w_j$, $i$-th eigenvalue and eigenstate are approximated by $\braket{\psi_i(\vec{\theta})|H|\psi_i(\vec{\theta})}$ and $\ket{\psi_i(\vec{\theta})}$, respectively.
Compared with the VQE for the ground state, evaluating the cost function of the SSVQE takes more computational cost (runs of quantum circuits and measurements) by $k$ times because one need to evaluate $\braket{\psi_i(\vec{\theta})|H|\psi_i(\vec{\theta})}$ for each $i=1,\ldots,k$ separately and combine them.
Moreover, the parameterized unitary circuit must be deeper to express the excited states because they are generally more entangled than the ground state. 
To implement the deeper unitary circuit, the fidelity required for the near-term quantum computers is tougher than that for the VQE, and more parameters need to be optimized so it will take longer time for the cost function to converge.

\section{Jordan-Wigner transformation\label{apdx:jw-transformation}}
The Jordan-Wigner transformation~\cite{Jordan1928} converts the fermionic creation and annihilation operators to the spin (qubit) operators faithfully preserving the algebra.
It regards the vacuum state $\ket{0}$ as the down spin $\ket{\downarrow}$ and the occupied state $\ket{1}$ as the up spin $\ket{\uparrow}$. 
The algebra of the fermionic operators follows the anti-commutation relations 
\begin{align}
    \{c_m, c_n^\dagger\} = \delta_{mn}, \qquad \{c_m, c_n\} = \{c_m^\dagger, c_n^\dagger\} = 0
\end{align}
where $c^\dagger_m$ and $c_m$ are the fermionic creation and annihilation operators acting on the $m$-th lattice site, respectively, and $\delta_{mn}$ is the Kronecker delta. 
These relations can be represented in terms of the spin operators if one replaces the fermionic operators as
\begin{align}
    c^\dagger_m \rightarrow (-1)^{m-1} Z_1 Z_2 \cdots Z_{m-1} \sigma^+_m, \qquad 
    c_m \rightarrow (-1)^{m-1} Z_1 Z_2 \cdots Z_{m-1} \sigma^-_m
\end{align}
where $\sigma^+_m = (X_m + iY_m)/2$ and $\sigma^-_m = (X_m - iY_m)/2$.

\section{Nonlinearity of excited-state properies\label{apdx:non-linearity}}
In this Appendix, we present further analysis of the 2-site Hubbard model discussed in Sec.~\ref{sec:discussion}.
The exact expressions of the excited-state properties of the Hubbard model in terms of the elements of the classical vector $\mathbf{x}_{\ket{\psi_0(r)}}$ present specific examples demonstrating that they may and may not be approximated with a linear model, given the classical vector. 

There are $4^4=256$ Pauli operators (from $I_0I_1I_2I_3$ to $Z_0Z_1Z_2Z_3$) which may act on the Hilbert space for the 2-site Hubbard model~(\ref{eq:Hubbard}).
Exhaustive search for all of these Pauli operators reveals that only two functions of $U$ appear as the ground state expectation values:
\begin{align}
 f_1(U) &:= \frac{1}{ \sqrt{1+\frac{16}{U^2}} } =-\braket{Z_0 Z_1}_\mrm{GS}=\braket{Z_0Z_3}_\mrm{GS}=\braket{Z_1Z_2}_\mrm{GS}=-\braket{Z_2Z_3}_\mrm{GS}\\ &=-\braket{X_0X_1X_2X_3}_\mrm{GS}= -\braket{X_0Y_1X_2Y_3}_\mrm{GS}=-\braket{Y_0X_1Y_2X_3}_\mrm{GS}=-\braket{Y_0Y_1Y_2Y_3}_\mrm{GS},  \\ 
  f_2(U) &:= \frac{1}{ \sqrt{1+\frac{U^2}{16}} } = \braket{X_0 Z_1 X_2}_\mrm{GS}=\braket{Y_0Z_1Y_2}_\mrm{GS}=\braket{X_1Z_2X_3}_\mrm{GS}=\braket{Y_1Z_2Y_3}_\mrm{GS} \\
  &=-\braket{X_0X_2Z_3}_\mrm{GS}=-\braket{Y_0Y_2Z_3}_\mrm{GS}=-\braket{Z_0X_1X_3}_\mrm{GS}=-\braket{Z_0Y_1Y_3}_\mrm{GS}.
\end{align}
In other words, for any choice of the entangler $U_\mrm{ent}$, all components of the classical vector $\mathbf{x}_{\ket{\psi_0(U)}} = \left( \braket{\psi_0(U)|U_\mrm{ent}^\dagger X_0 U_\mrm{ent}|\psi_0(U)}, \cdots, \braket{\psi_0(U)|U_\mrm{ent}^\dagger Z_{N-1} U_\mrm{ent}|\psi_0(U)} \right)^T$ will be written as a linear combination of $f_1(U)$ and $f_2(U)$.

Now, we can see that the nonlinearity in the classical unit is not necessary for a small value of $U$, but it is for a large $U$.
For $0 < U \ll 1$, $f_1(U) \approx U/4$ and $f_2(U) \approx 1$, and the excited-state properties in terms of these functions are
\begin{align}
\Delta E_1  \approx 2 - \frac{U}{2}, \quad 
\Delta E_2 \approx 2 + \frac{U}{2}, \quad
\|\bm{\mu}_\mrm{eg}\| \approx \frac{1}{\sqrt{2}} - \frac{U}{8\sqrt{2}},
\end{align}
ignoring $\mathcal{O}(U^2)$ terms.
In this case, the excited-state properties can be expressed easily as the linear combination of $f_1(U)$ and $f_2(U)$.
On the other hand, when $U$ is large, i.e. when $0<1/U\ll 1$, it follows that $f_1(U) \approx 1$ and $f_2(U) \approx 4/U$, and the excited-state properties can be expressed as
\begin{align}
\Delta E_1  \approx \frac{4}{U},  \quad
\Delta E_2 \approx  U, \quad
\|\bm{\mu}_\mrm{eg}\| \approx \frac{2}{U},
\end{align}
ignoring $O(1/U^2)$ terms.
$\Delta E_2$ may not be expressed as a linear combination of $f_1$ and $f_2$. 

\begin{figure*}
    \includegraphics[width=.42\textwidth]{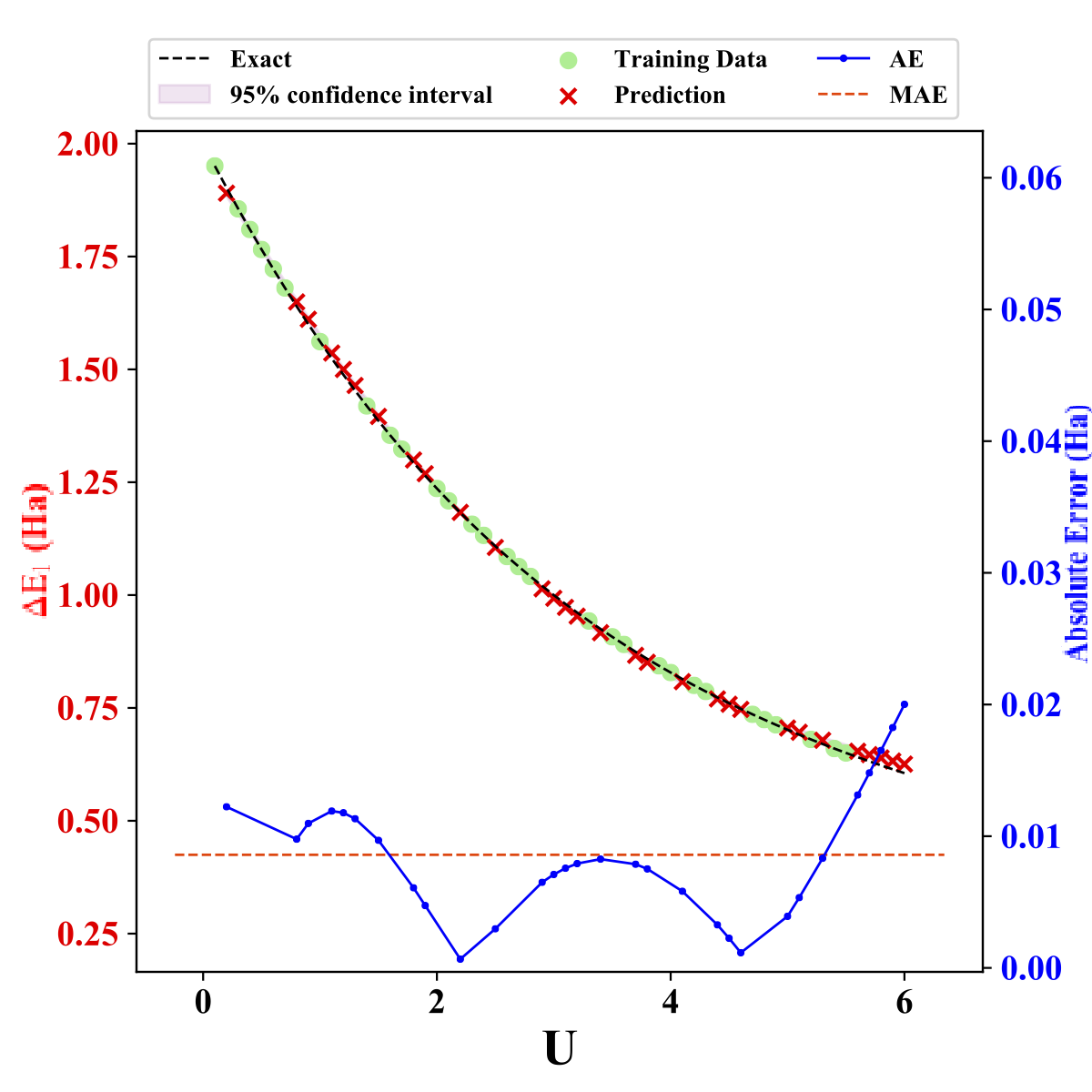} 
    \includegraphics[width=.42\textwidth]{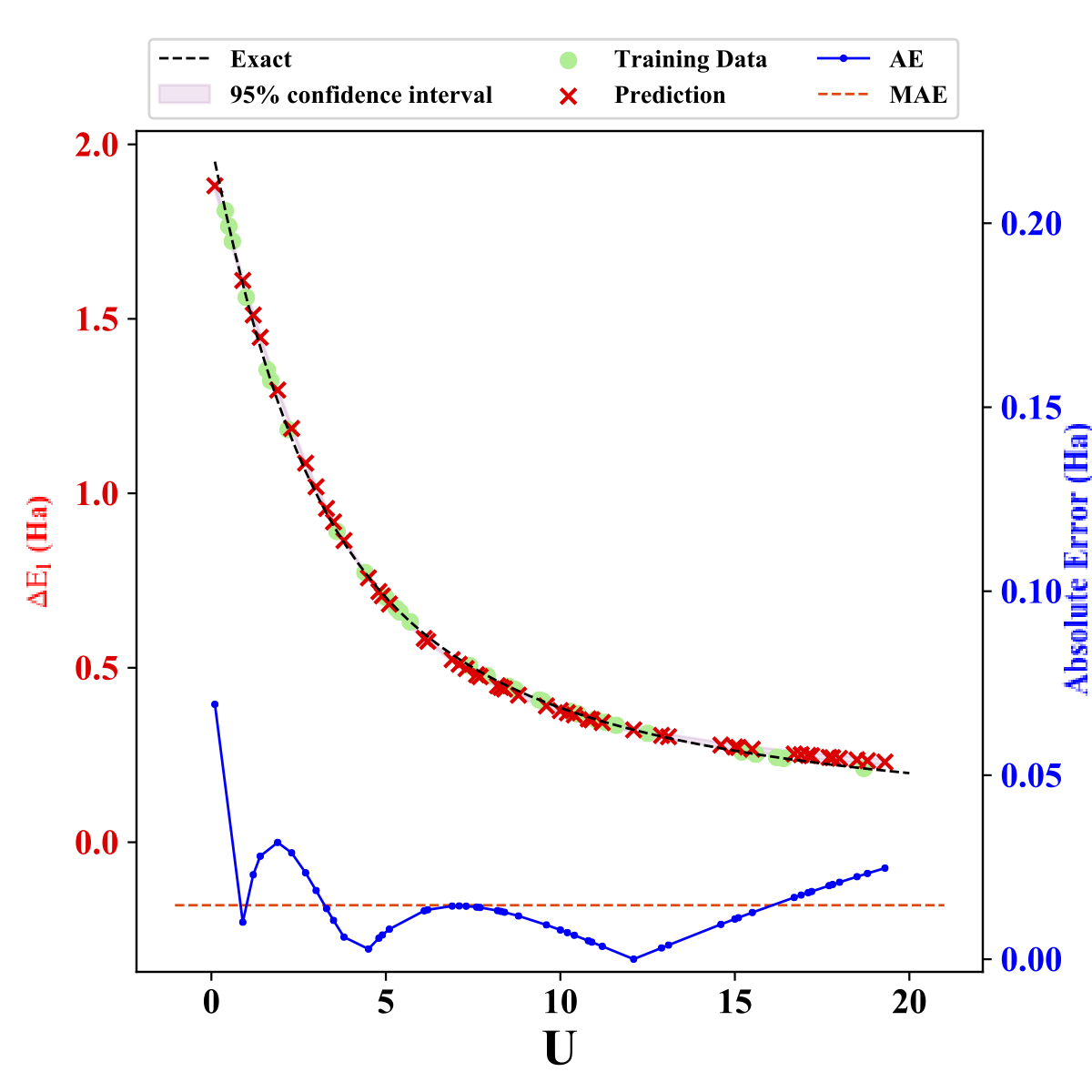} \\
    \includegraphics[width=.42\textwidth]{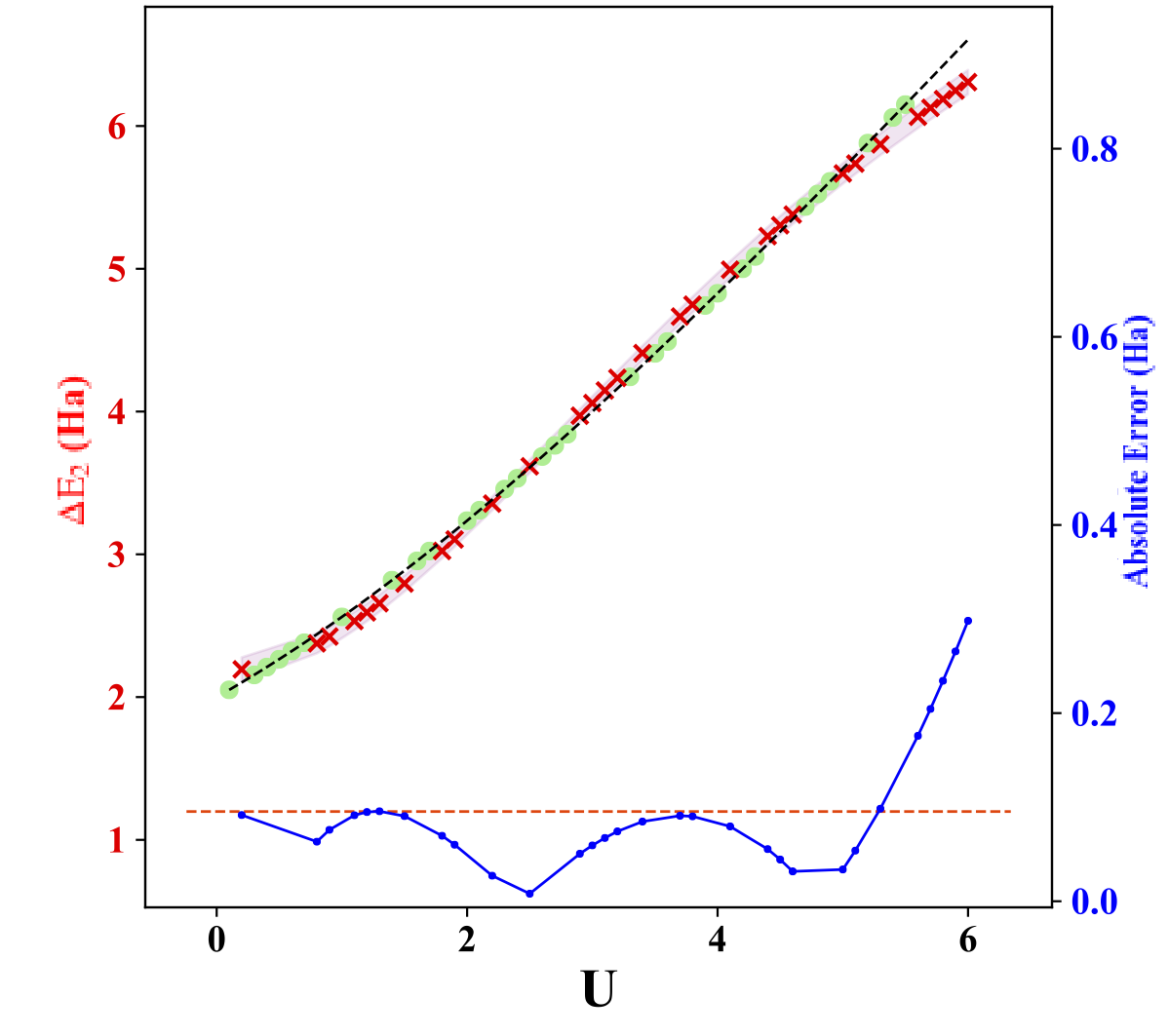} 
    \includegraphics[width=.42\textwidth]{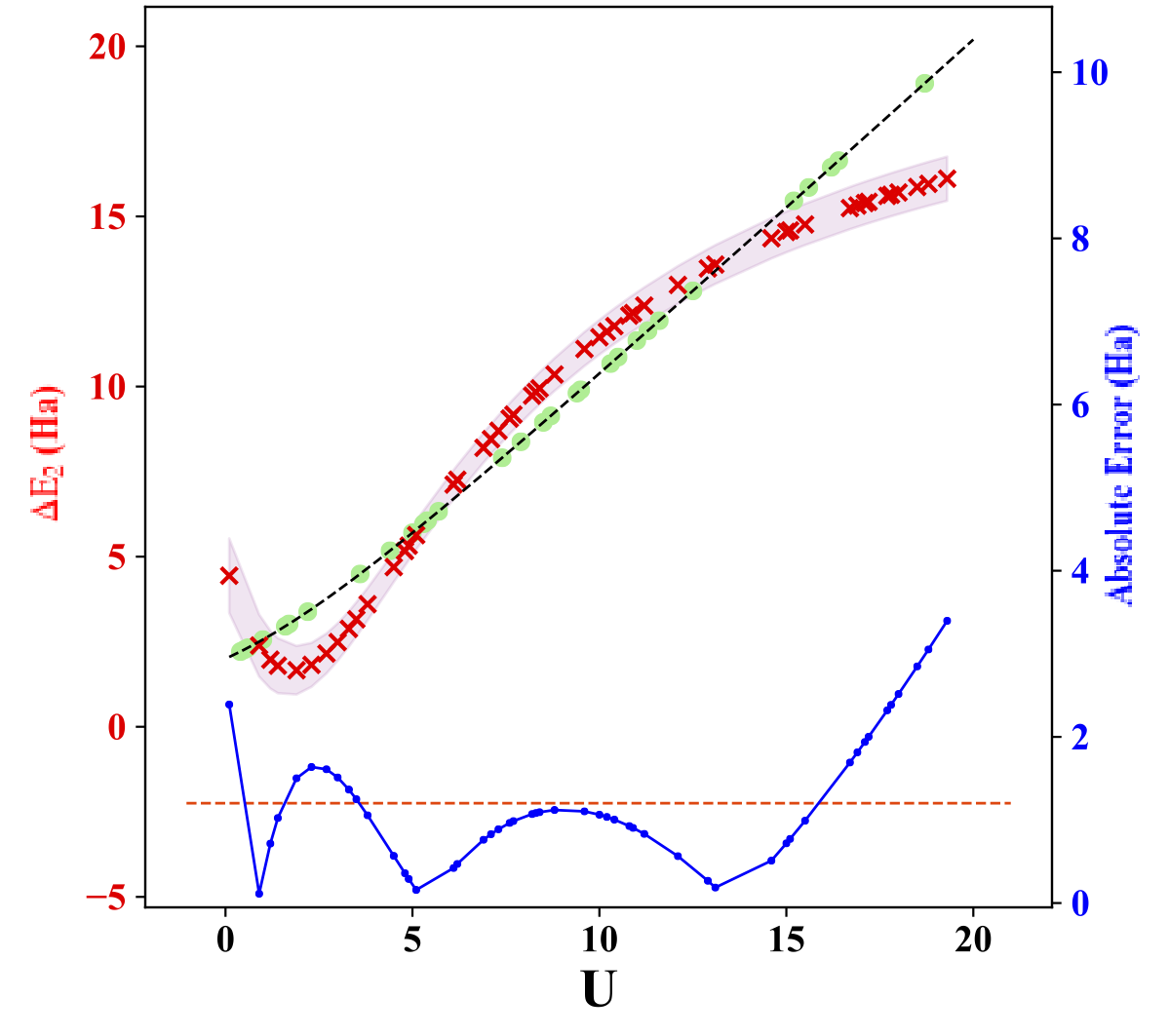} \\ 
    \includegraphics[width=.42\textwidth]{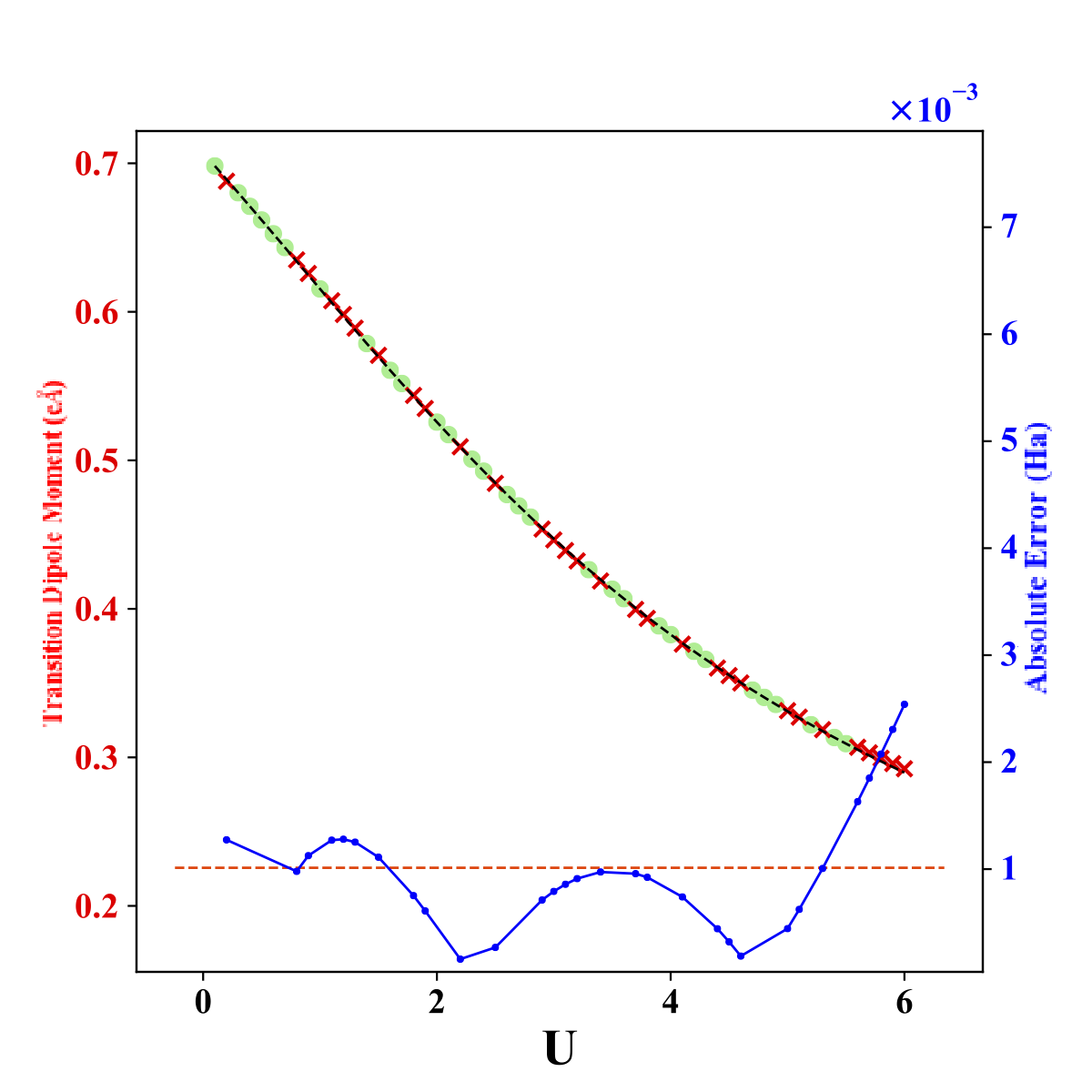}  
    \includegraphics[width=.42\textwidth]{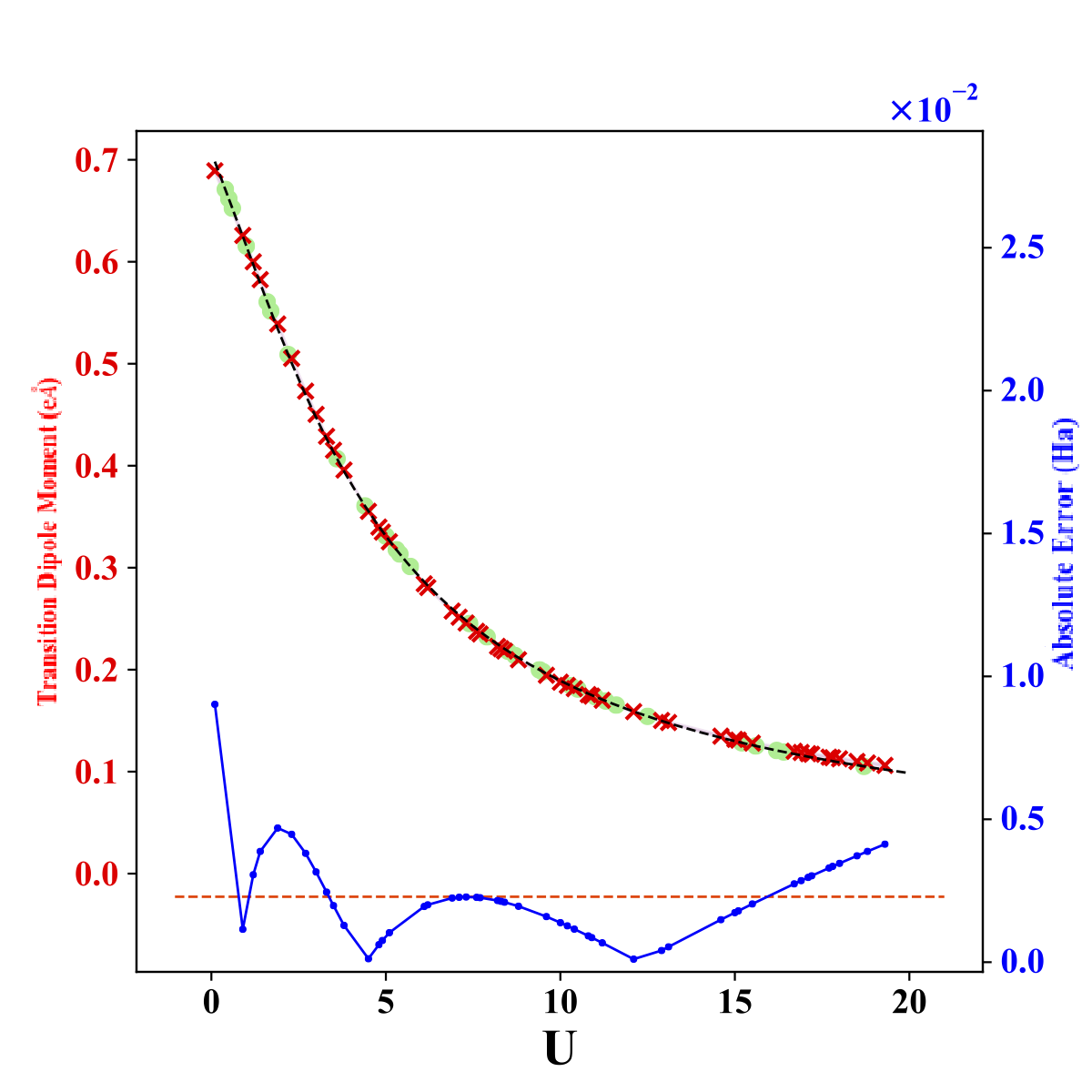}
     \caption{\label{fig:Hubbard}
     The  prediction  results  by  the  trained  model for the 2-site Hubbard model~(\ref{eq:Hubbard}) with two datasets sampled from (Case 1) $U \in [0.1, 6]$ (left column) and (Case 2) $U \in [0.1, 20]$ (right column).
     Top, middle, and bottom panels display the first, second excitation energies and the transition amplitude, respectively.
     The blue circles indicate the training data points and the red crosses do the predictions. The exact values are displayed as the black line. 
     Those values are read from the left ticks of each panel. 
     The blue plot lines with circles represent the absolute errors between the predictions and the exact values, and the orange line indicates their mean. 
     These error values are read from the right ticks of each panel.}
\end{figure*}

To support the observation, we also perform a numerical simulation for the 2-site Hubbard model.
We randomly sample 30 distinct values of $U$ for the training data and 50 distinct values of it for the test data in the range of $U \in [0.1, 6]$ (Case 1) and $U \in [0.1, 20]$ (Case 2).
The ground state wavefunction of $H_\mrm{Hub}(U)$ is prepared by the exact diagonalization.
Instead of using an entangler, here we define the classical vector $\mathbf{x}_{\ket{\psi_0(U)}}$ as $(\braket{Z_0 Z_1}_\mrm{GS}, \braket{X_0 Z_1 X_2}_\mrm{GS})^T$.
The linear regression to learn the excited-state properties $\mathbf{y} = (\Delta E_1, \Delta E_2, \|\bm{\mu}_\mrm{eg}\|)^T$ from $\mathbf{x}_{\ket{\psi_0(U)}}$ is performed both for Case 1 and Case 2.
All values of $\mathbf{x}_{\ket{\psi_0(U)}}$ and $\mathbf{y}$ are standardized by using the mean and the standard deviation of the training dataset respectively during the training process of the LR.
The results are shown in Figure~\ref{fig:Hubbard}.
The LR predicts the excites state properties almost perfectly for small values of $U$ as one may see in the results for Case 1, whereas it fails once one tries to learn and predict from the data with large $U$ values as shown in the results for Case 2, especially evident from the prediction of $\Delta E_2$.
Those results support our expectation that a linear classical unit can sufficiently approximate the excited-state properties in the case of small $U$, but it may not for a large $U$.
We consider a similar mechanism applies to the numerical simulations of small molecules in Sec.~\ref{sec:result},
where atomic-spacings are not very small so that the Coulomb repulsion $U$ is not large.

\section{Time evolution operator as the entangler: its action on the single qubit operators \label{apdx:entangler-analysis}}
In the numerical simulations in Sec.~\ref{sec:result}, we adopt the time evolution operator $e^{-iH_\mrm{rand}T}$, where $H_\mrm{rand}$ is a random Hamiltonian and $T$ is a fixed time for the evolution, as the entangler $U_\mrm{ent}$.
In this case, we can intuitively understand the effect of the entangler by considering the time evolution of the single-qubit Pauli operators $\bigcup_{i=0}^{N-1}\{X_i, Y_i, Z_i\}$ in the Heisenberg picture, where $N$ is the number of qubits in the system.

As discussed in Section~\ref{sec:model-desc}, the information we obtain as the outputs of the quantum circuit in our model are the expectation values of the complicated operators $\bigcup_{i=0}^{N-1}\{U_\mrm{ent}^\dag X_i U_\mrm{ent}, U_\mrm{ent}^\dag Y_i U_\mrm{ent}, U_\mrm{ent}^\dag Z_i U_\mrm{ent}\}$ for the ground state $\ket{\psi}$.
If we expand the random Hamiltonian as $H_\mrm{rand} = \sum_{P\in \{I, X, Y, Z\}^{\otimes N}} h_P P$ in the basis of $N$-qubit Pauli operators $P$ and coefficients $h_P$, it follows
\begin{align}
  U_\mrm{ent}^\dag X_i U_\mrm{ent} = e^{iH_\mrm{rand}T} X_i e^{-iH_\mrm{rand}T}
  = X_i + iT\sum_{P\in \{I, X, Y, Z\}^{\otimes N}, P \neq I} h_P [P, X_i] + O(T^2),
\end{align}
where $[A, B] = AB - BA$.
Same for the $Y_i$ and $Z_i$ operators.
$i[P, X_i]$ is another $N$-qubit Pauli operator with larger support (i.e. the number of qubits on which $i[P, X_i]$ acts nontrivially is larger) than $X_i$, if $P$ nontrivially acts on the $i$-th qubit and one or more other qubits. 
As seen in Section~\ref{subsec:hubbard} and \ref{apdx:non-linearity}, to estimate the excited-state properties, we generally need the information of the expectation values of certain Pauli operators nontrivially acting on multiple qubits. 
Hence, if the set of the Pauli operators $\{i[P, X_i]\}_P$ includes such required operators, the machine learning unit may automatically find
them and construct the excited-state properties as a function of the expectation values. 
We note that $O(T^2)$ terms contain the terms like $[[P', [P, X_i]]$, $O(T^3)$ terms contain the terms like $[[P'', [P', [P, X_i]]]$, and so on, so even when $\{i[P, X_i]\}_P$ does not contain the required operators, they may be contained in these higher-order terms, and the machine learning unit may find them if $T$ is large enough so that the higher-order terms in $T$ contribute enough to $U_\mrm{ent}^\dag X_i U_\mrm{ent}$.
Hence, the operators $\{U_\mrm{ent}^\dag X_i U_\mrm{ent}, U_\mrm{ent}^\dag Y_i U_\mrm{ent}, U_\mrm{ent}^\dag Z_i U_\mrm{ent}\}_{i=0}^{N-1}$ are constituted from more long-ranged, multi-qubit Pauli operators if (1) the random Hamiltonian $H_\mrm{rand}$ contains stronger and longer-ranged interactions and/or (2) the time for the evolution becomes larger.
This means that the expectation values of $\bigcup_{i=0}^{N-1}\{U_\mrm{ent}^\dag X_i U_\mrm{ent}, U_\mrm{ent}^\dag Y_i U_\mrm{ent}, U_\mrm{ent}^\dag Z_i U_\mrm{ent}\}$ for the ground state $\ket{\psi}$ bring more information of $\ket{\psi}$ and the original molecular Hamiltonian than those of $\bigcup_{i=0}^{N-1}\{X_i, Y_i, Z_i\}$ for the ground state, 
and there is more chance for the machine learning unit to successfully predict the excited-state properties from the information. 
We note that the physical picture of the spreading of the single-qubit Pauli operators over the whole system under chaotic Hamiltonians was discussed in Ref.~\cite{Ho2017}.

\section{Dependence of the excited-state prediction on the number of shots for the VQE \label{apdx:vqe-shots}}
\begin{figure*}
    \centering
    \includegraphics[width=0.45\linewidth]{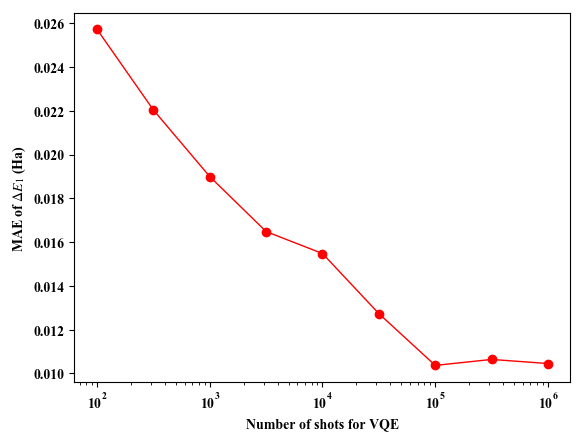}
    \includegraphics[width=0.45\linewidth]{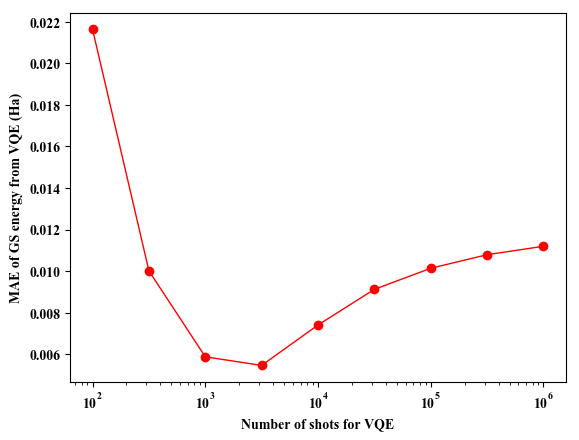}
    \caption{\label{fig:mae-vs-vqe}(Left) Dependence of the MAEs of the predictions of $\Delta E_1$ for LiH on the number of shots employed in performing the VQE.
    (Right) Dependence of the MAEs between the ground state energy obtained by the VQE and the exact one computed by diagonalization of the molecular Hamiltonian of LiH on the number of shots to perform the VQE.
    }
    
\end{figure*}

In this appendix, we present how the accuracy of the predictions from our model varies with the number of shots used to perform the VQE for preparing the dataset of the ground states.
We carried out the same numerical simulation for the LiH molecules in the noisy situation as described in Sections~\ref{subsec:simulation_def} and \ref{subsubsec:noisy_result}, but with the various numbers of shots ranging from 100 to $10^6$ for the computation of the VQE, instead of fixing it to 10000 shots. 
Left panel of Figure~\ref{fig:mae-vs-vqe} displays the MAE for the predictions of $\Delta E_1$ versus the number of shots for the VQE, showing that the MAE decreases almost monotonically with the number of shots. 
Right panel of Figure~\ref{fig:mae-vs-vqe} shows the MAE between the ground state energy computed by the VQE and the exact one obtained by diagonalization of the molecular Hamiltonians of LiH, as a function of the number of shots. 
Interestingly, the accuracy of the VQE has an empirical overhead at around $10^3-10^4$ shots and gradually saturates the infinite-shots limit which is non-zero because of the presence of the noise.
Two panels of Figure~\ref{fig:mae-vs-vqe} suggest that the accuracy of the predictions of the excited-state properties is almost independent of the precision of the computation result from the VQE. Rather, it depends on the number of shots, and one may simply increase the shots to obtain estimations with higher accuracy. 
We leave a deeper analysis of this curious dependence as future work.

\section*{References}
\bibliography{bibliography}

\end{document}